\documentclass{aa}
\begin{document}
\title{Trends in torques acting on the star during a star-disk magnetospheric
interaction}

\titlerunning{Trends in torques acting on a star}
\authorrunning{\v{C}emelji\'{c} \& Brun}
\author{M. \v{C}emelji\'{c}
\inst{1,2,3}
\and
A.S. Brun
\inst{4}
}
\offprints{M. \v{C}emelji\'{c}}
\institute{Research Centre for Computational Physics and Data Processing,
Institute of Physics, Silesian University in Opava, Bezru\v{c}ovo
n\'{a}m. 13, CZ-746 01 Opava, Czech Republic, orcid-id: 0000-0002-3434-3621,
\email{miki@camk.edu.pl}
\and
Nicolaus Copernicus Astronomical Center, Bartycka 18, 00-716
Warsaw, Poland
\and
Academia Sinica, Institute of Astronomy and Astrophysics, P.O. Box 23-141,
Taipei 106, Taiwan
\and
 D\'epartement d'Astrophysique/AIM, CEA/IRFU, CNRS/INSU, Univ.
Paris-Saclay \& Univ. de Paris, 91191 Gif-sur-Yvette, France,
orcid-id: 0000-0002-1729-8267, \email{sacha.brun@cea.fr}
} 

\date{Received ??; accepted ??}
\abstract
{}
{We assess the modification of angular momentum transport in various
configurations of star-disk accreting systems based on numerical
simulations with different parameters. In particular, we quantify the
torques exerted on a star by the various components of the flow and field
in our simulations of a star-disk magnetospheric interaction.
}
{In a suite of resistive and viscous numerical simulations, we obtained
results using different stellar rotation rates, dipole magnetic field
strengths, and resistivities. We probed a part of the parameter space
with slowly rotating central objects, up to 20\%\ of the Keplerian
rotation rate at the equator. Different components of the flow in
star-disk magnetospheric interaction were considered in the study:
a magnetospheric wind (i.e., the ``stellar wind'') ejected outwards from the
stellar vicinity, matter infalling onto the star through the accretion
column, and a magnetospheric ejection launched from the magnetosphere. We also
took account of trends in the total torque in the system and in each
component individually.
}
{We find that for all the stellar magnetic field strengths, B$_\star$, the anchoring radius
of the stellar magnetic field in the disk is extended with increasing
disk resistivity. The torque exerted on the star is independent of the stellar
rotation rate, $\Omega_\star$, in all the cases without magnetospheric ejections. In cases
where such ejections are present, there is a weak dependence of the anchoring
radius on the stellar rotation rate, with both the total torque in the system and
torque on the star from the ejection and infall from the disk onto the star
proportional to $\Omega_\star B^3$. The torque from a magnetospheric ejection is
proportional to $\Omega_\star^4$. Without the magnetospheric ejection, the
spin-up of the star switches to spin-down in cases involving a larger stellar
field and faster stellar rotation. The critical value for this switch is about 10\%\ of the Keplerian rotation rate.
}
   {}


\maketitle

\nolinenumbers
\section{Introduction}
In young stellar objects (YSOs) such as young T-Tauri stars, the process of
accretion of material from the initial cloud is almost complete and the
star is just about to start burning its thermonuclear fuel. During the gravitational
infall of matter onto a central object, an accretion disk is formed,
through which angular momentum is transported away from the central object. 

To construct consistent stellar spin-down models of the formation of Sun-like stars,
a magnetic field is required \citep{BouvierCebron15, Ahuir20}. Apart from the
stellar wind properties, a self-consistent treatment of star-disk
magnetospheric interaction need to be included in the considerations.
Analytical solutions for magnetic thin accretion disks are impossible
without imposing severe approximations because the system of equations is
not closed \citep{ckp19b}. This leaves us with self-consistent numerical
simulations as the only way to obtain a solution. Important steps have been undertaken by
\cite{ghl79a,ghl79b}, who found that to correctly describe the star-disk
magnetospheric interaction, it is necessary to include the rotating stellar
surface and corona as well as to extend the computational domain beyond the position of the corotation radius.
\cite{Matt05,Matt12a} discussed spin-down models and the dependence of
torque on stellar field stellar rotation rates. In a series of works,
\cite{R09,R13} investigated the star-disk interaction in numerical simulations,
as did \cite{ZF09,ZF13}, \cite{cem19}. These authors identified the typical geometry in simulations, with the magnetospheric wind, disk accretion flow, accretion column onto the central object, and (in cases where there is less magnetic diffusion in the disk) magnetospheric ejections.

Global numerical solutions describing the effect of magnetic field (in
particular, resistivity) on the transport of angular momentum in the
star-disk magnetosphere are still overly dependent on the chosen parameters
in the disk. The torques caused by the stellar wind and ejection are not enough
to counterbalance the torque exerted onto the central object by the material
accreted through the disk. Also, the influence of magnetic reconnection in
the disk corona remains undetermined.

There has not yet been any numerical simulations study that would extensively relate
the quantities of (turbulent) resistive MHD in the disk and star-disk
magnetospheric interaction as we do here. For comparisons with the observational
data, a study in a large part of parameter space is needed. Such a study
will yield trends; namely: expressions that are proportional to the dependence of
various components of torques on density, velocity, and magnetic field
components, as well as on the effective (anomalous) coefficients of both
viscosity and resistivity. 

Prescriptions from models, such as those given in \cite{GalZanAma19},
 match the observed quantities reasonably well, but they do depend on the
chosen limits of validity for each phase of pre-stellar evolution and
require unrealistic large stellar fields to counterbalance the disk
accretion for the spinning-up a star. Alternatively, a large mass loads in the
stellar wind of up to 10\% accretion rate are needed or, otherwise, an interplay
of the lighter stellar wind with the disk truncation near the corotation
radius. Consequently, more realistic models, informed by relevant
numerical simulations, are needed to remove the implicit model
requirements.

A comparison of the results from our numerical simulations with the
results from models as given in \cite{GalZanAma19} can serve both as a
guide in evaluation of the simulations results and as information for
the further refinements of the model. Following \cite{ZF09}, in \cite{cem19} we set a \citet[hereafter KK00]{KK00}
disk as the initial condition in our simulations, adding the stellar
dipole magnetic field. We obtained an ``atlas'' of magnetic solutions, with
three types of solutions for the slowly rotating star: with and without the
accretion column, and with a magnetospheric ejection above the accretion
column. An initial parameter study was performed in the 64 cases with
different magnetic field strengths, stellar rotation rates and (anomalous)
resistivity parameters. We found continuous trends in the average angular
momentum flux transported onto the stellar surface through the accretion
column from the disk onto the star. We also found a trend in the angular
momentum flux expelled from the system in the magnetospheric ejection,
which forms in the solutions with the resistive coefficient
$\alpha_{\mathrm m}=0.1$ in our simulations. Here, we describe the subsequent
analysis of our 64 runs.

In the present study, contributions in the accretion flow onto the star
are decomposed by the torque they exert on the stellar surface. Any speeding
up or slowing down of the stellar rotation depends precisely upon the
distance separating the field lines from the star to their footpoint
anchoring in the disk. We also indicate trends in the torque exerted on
a star by the magnetospheric ejection and stellar wind, and we check for
trends in the total torque in the system with respect to the stellar
magnetic field, stellar rotation rate, and resistivity.

In Sect.~\ref{short}, we briefly present our numerical setup and give a general
overview of the obtained quasi-stationary states and flows. Trends in the
reach of stellar magnetic field in the disk are outlined in Sect.~\ref{reach}.
In Sect.~\ref{angmom}, we present the details of our computation of torques
from the different components in the star-disk system and present
the obtained trends in Sect.~\ref{trends}. We give our conclusions in Sec.~\ref{conclu}.

\section{Short overview of simulations}\label{short}
\begin{figure}
\includegraphics[width=\columnwidth]{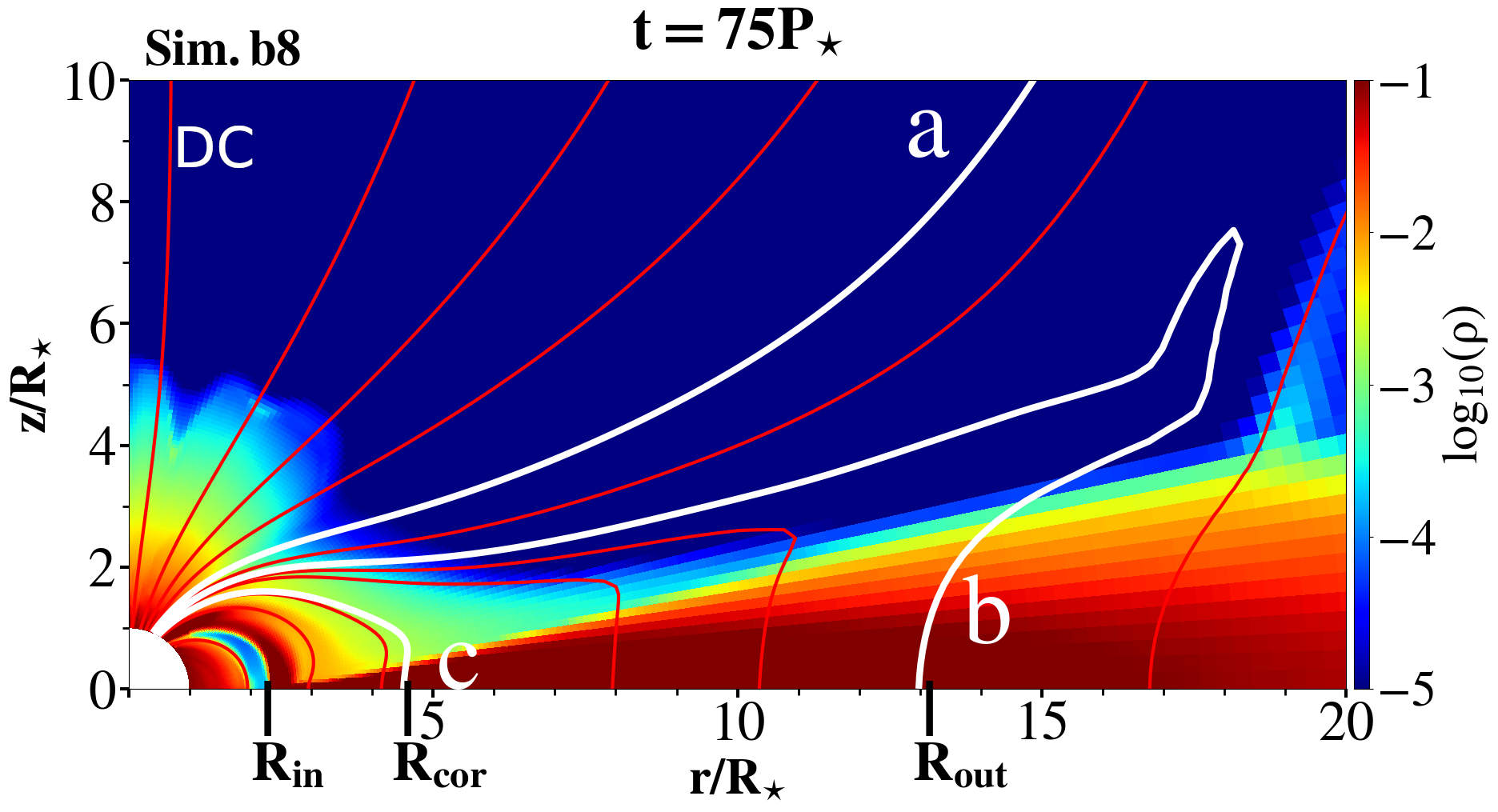}
\includegraphics[width=\columnwidth]{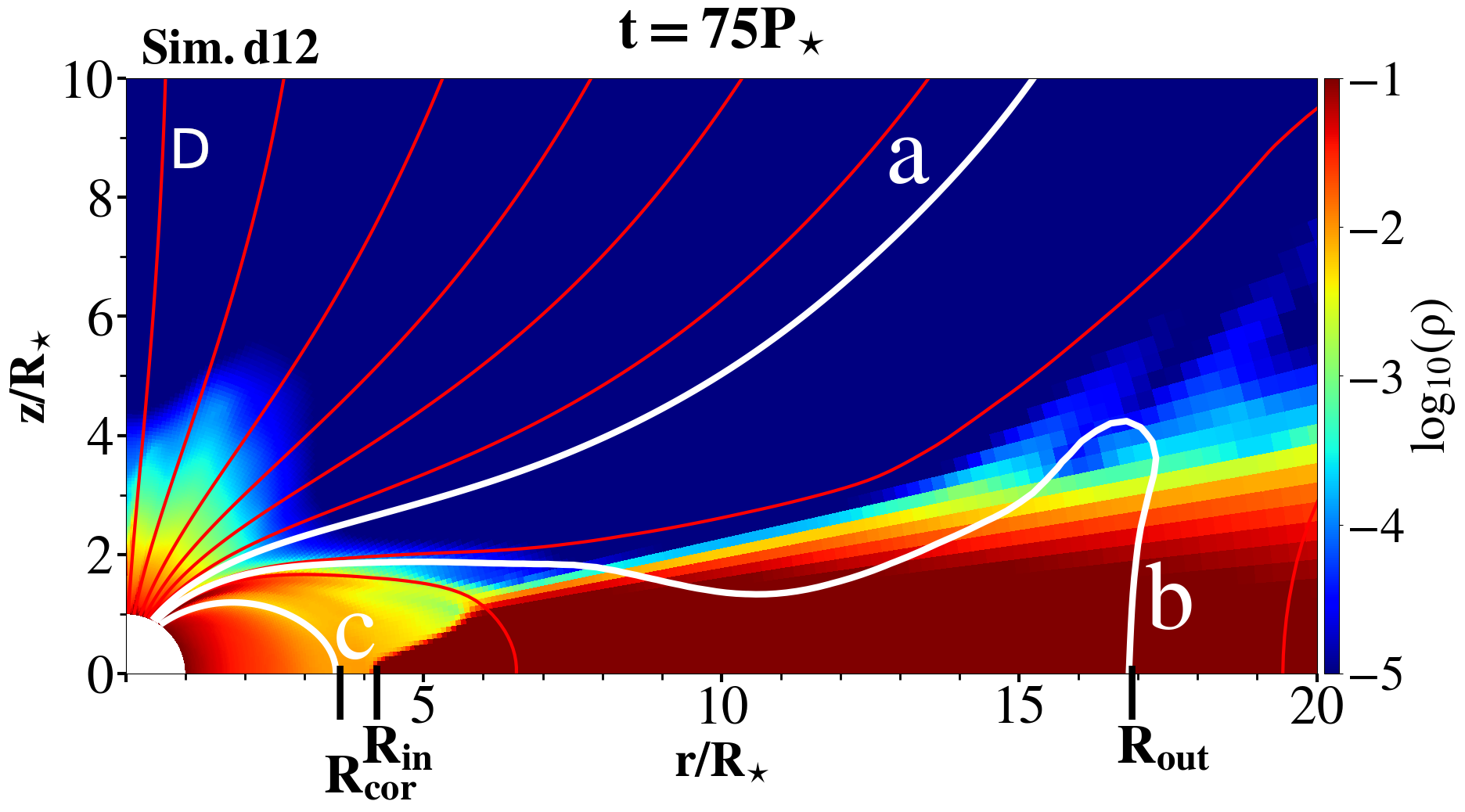}
\includegraphics[width=\columnwidth]{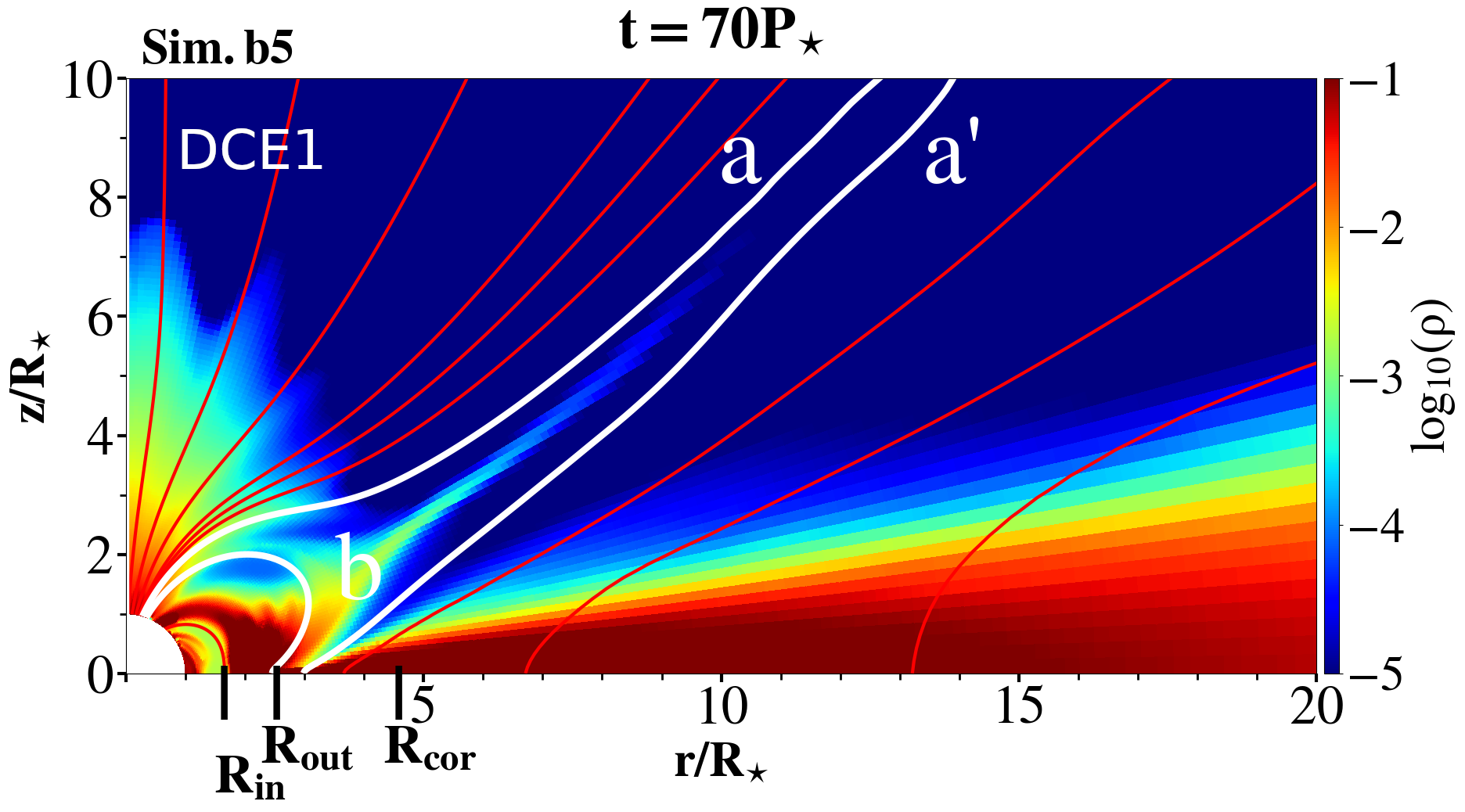}
\includegraphics[width=\columnwidth]{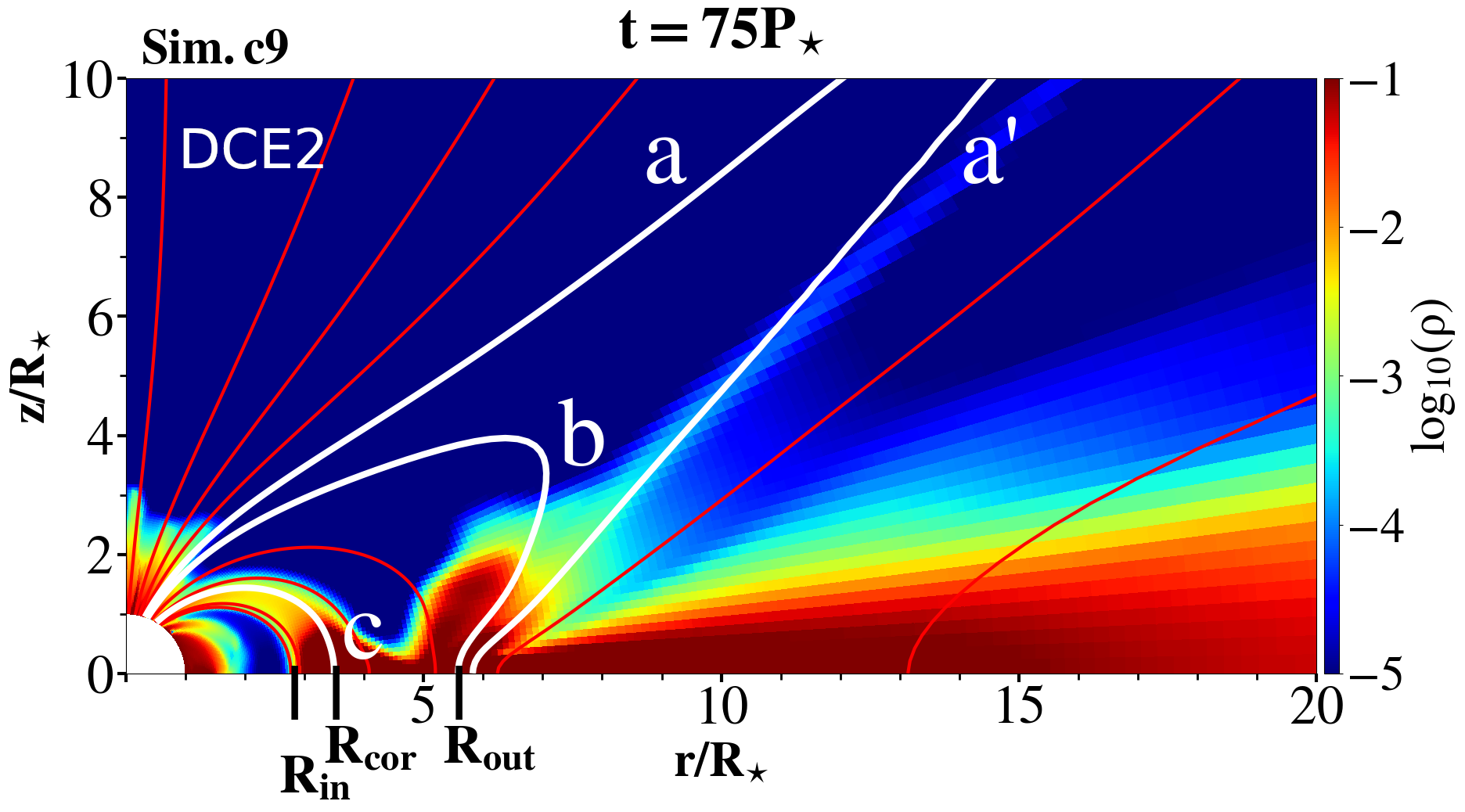}
\caption{Four typical geometries in our solutions: Disk$+$Column (DC),
Disk (D), Disk$+$Column$+$Ejection1 (DCE1), and Disk$+$Column$+$Ejection2 (DCE2) illustrated by snapshots in quasi-stationary states in simulations
b8, d12, b5, and c9 (top to bottom, respectively).
We show the density
in a logarithmic color grading, with a sample of magnetic field lines
in red solid lines. White lines labeled {\bf a}, {\bf a'}, {\bf b,} and
{\bf c} delimit the components of the flow between which we integrate
the fluxes; $R_{\mathrm in}$, $R_{\mathrm out}$ and $R_{\mathrm cor}$
are the inner disk edge, the furthest outer reach of closed stellar
field in the disk and the corotation radius, respectively. 
}
\label{fig:sols1}
\end{figure}
\begin{figure}
\includegraphics[width=0.49\columnwidth]{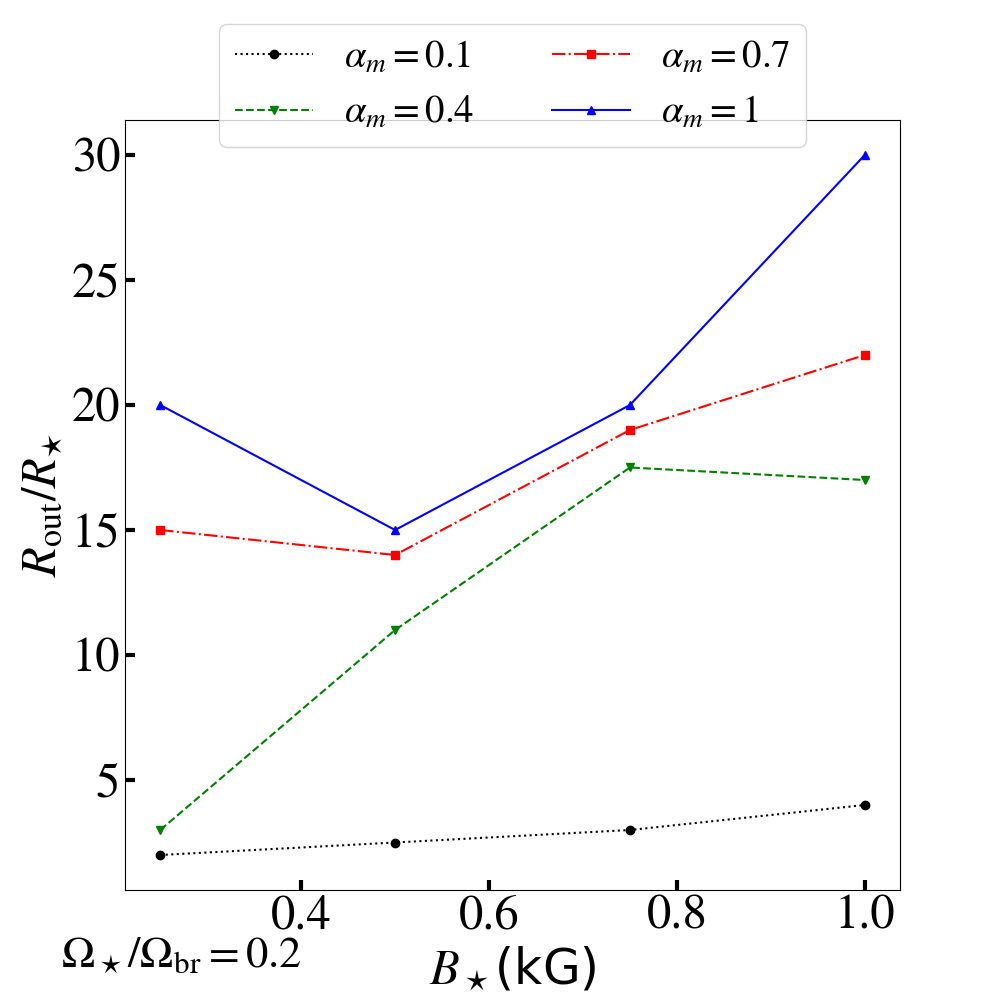}
\includegraphics[width=0.49\columnwidth]{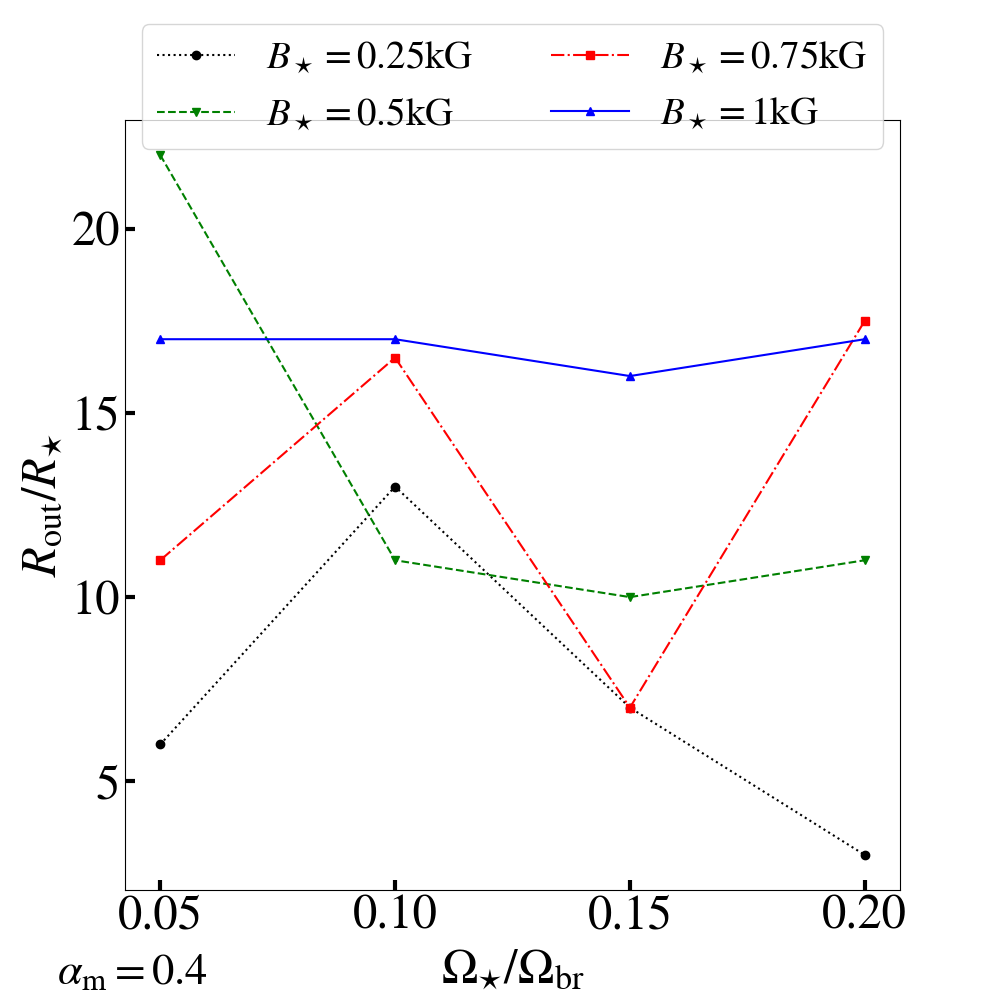}
\includegraphics[width=0.49\columnwidth]{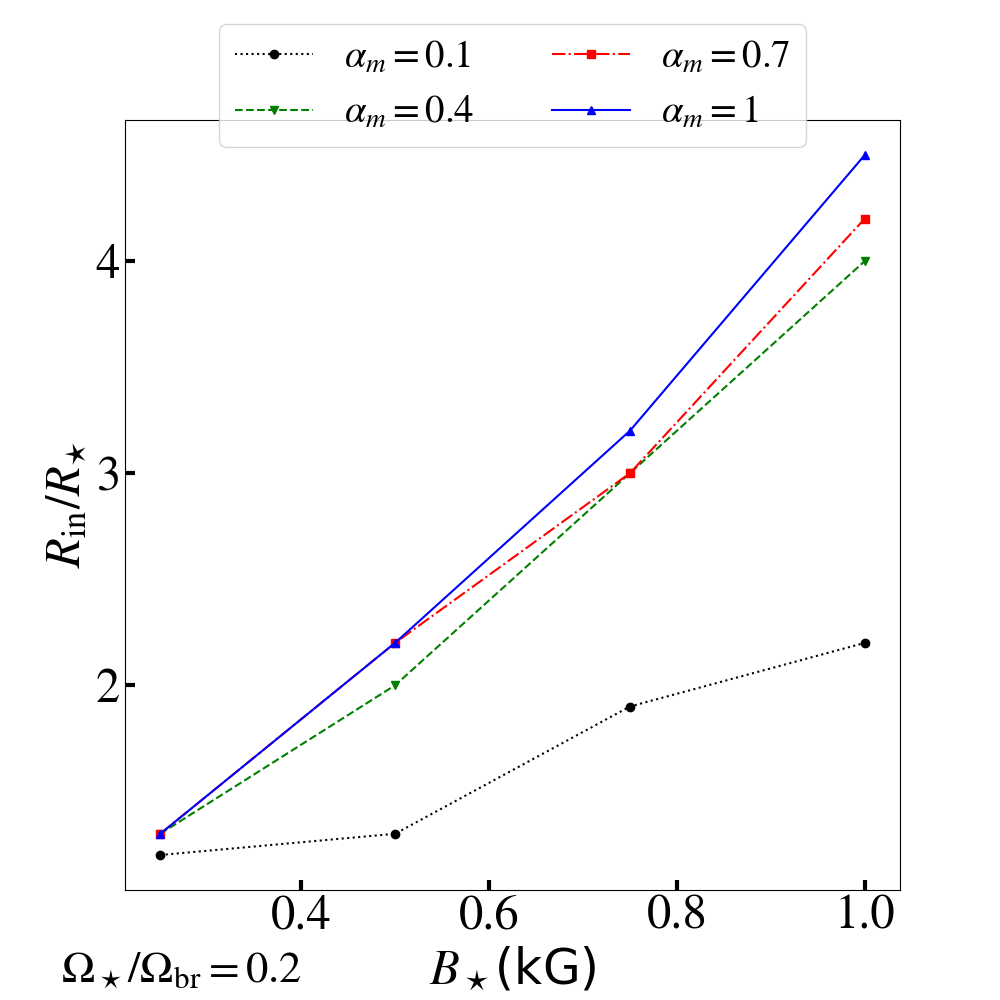}
\includegraphics[width=0.49\columnwidth]{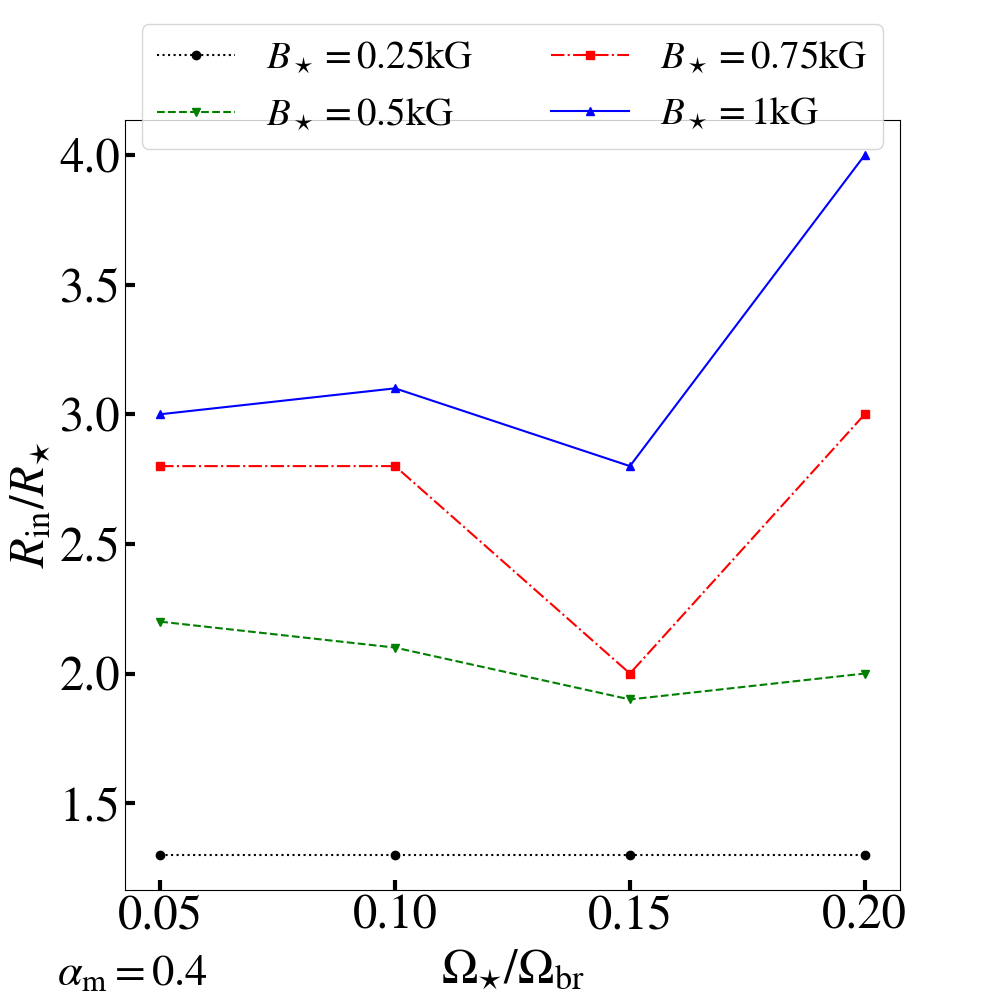}
\caption{Positions of the anchoring radius, $R_{\mathrm out}$, for
the furthest stellar magnetic field line in the disk that still reaches the star and the inner disk radius, $R_{\mathrm in}$.
{\it Top panels}: $R_{\mathrm out}$ increases
with the increasing resistive coefficient $\alpha_{\mathrm m}$ for
all stellar magnetic fields (left) and position of the
$R_{\mathrm out}$ as a function of stellar rotation rate in the cases
without magnetospheric ejection shows a decreasing trend with the stellar magnetic field (right). It is only in the cases with $B_\star=0.5~kG$ that we obtain a departure from this trend, with the largest $R_{\mathrm out}$ at the slowest stellar rotation rate.
{\it Bottom panels}: Inner disk radius, $R_{\mathrm in}$, is increasing
in the majority of the cases with same parameters as in the above panels
for $R_{\mathrm out}$.
}
\label{routrad}
\end{figure}
\begin{figure}
\includegraphics[width=0.49\columnwidth]{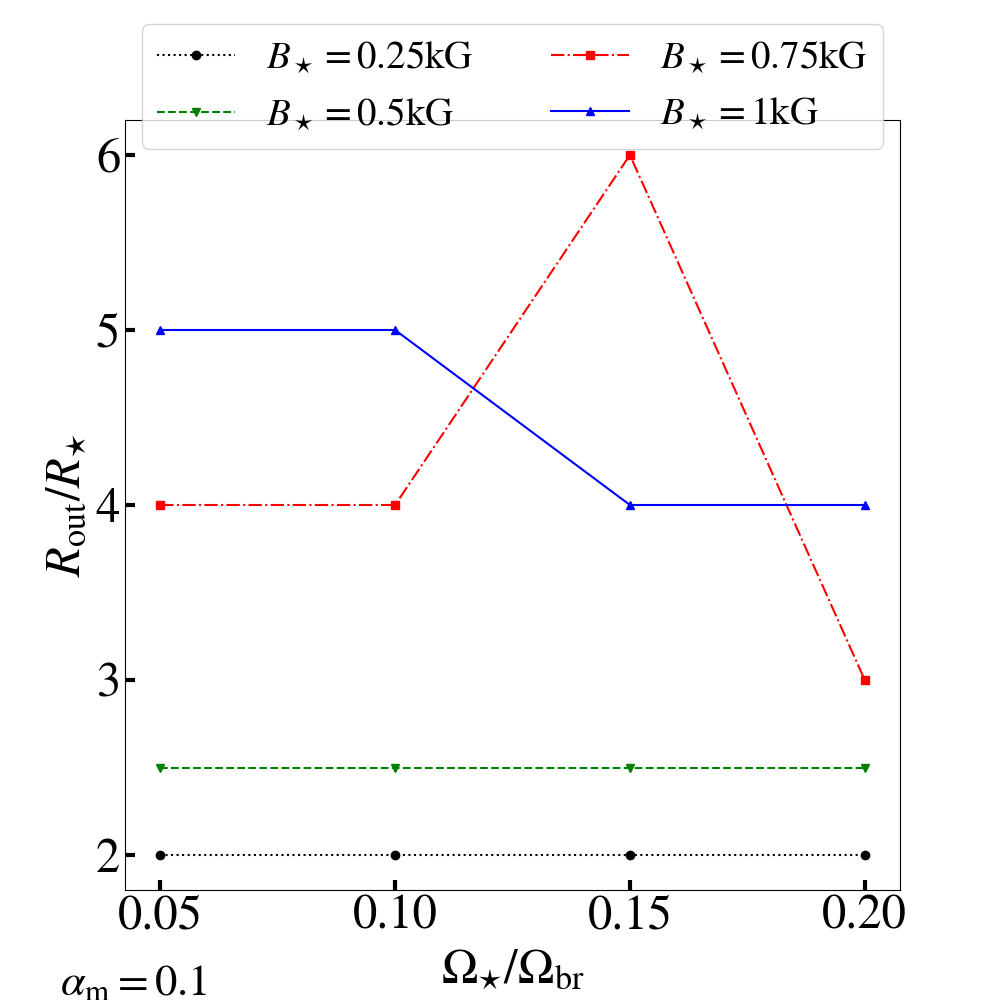}
\includegraphics[width=0.49\columnwidth]{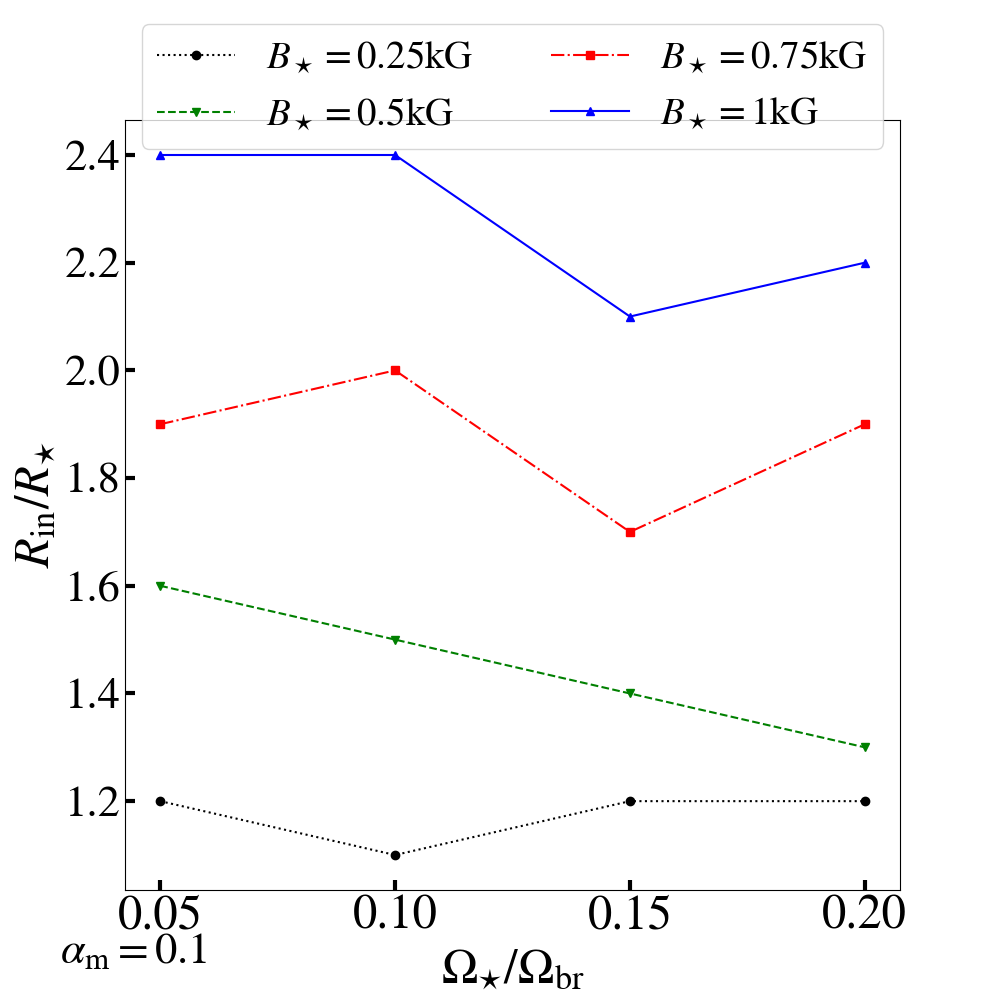}
\caption{Variation of the furthest anchoring radius of the
stellar field in the disk $R_{\mathrm out}$ with the stellar field in
the cases with $\alpha_{\mathrm m}=0.1$ (which are all with a magnetospheric
ejection launched from the system) is not large ({\it left)}. Without the magnetospheric
ejection, the scattering in this result is much larger. Inner disk radius $R_{\mathrm in}$ in the same cases displays a clear
increasing trend with the increase of stellar magnetic field ({\it right)}.
}
\label{routradb}
\end{figure}
\begin{figure}
\includegraphics[width=0.9\columnwidth]{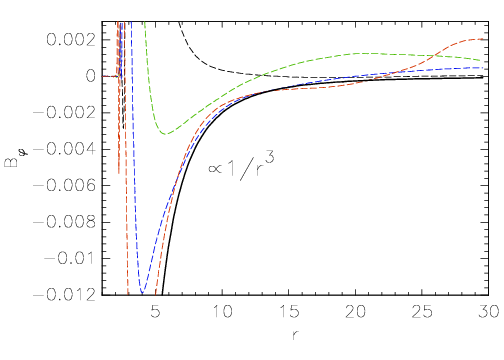}
\includegraphics[width=0.9\columnwidth]{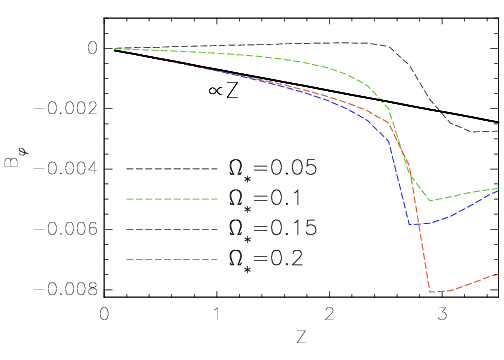}
\caption{Toroidal magnetic field values along the disk surface
({\it top}) and along the disk height (measured from the
disk equatorial plane) at $R=12R_\star$ ({\it bottom})
are shown in terms of their dependence of stellar rotation rates. The
values are averaged over 10 stellar rotation periods
during the quasi-stationary states, in the cases
with $\alpha_{\mathrm m}$=1 and $B_\star=500$~G (simulations
b1, b5, b9, and b13). In black, green, blue, and red dashed lines we display the values for 0.05, 0.1, 0.15, and 0.2 of
$\Omega_{\mathrm br}$, respectively. The black solid lines
provide the reference for typical radial and vertical dependence.}
\label{bphis}
\end{figure}
\begin{figure}
\includegraphics[width=\columnwidth]{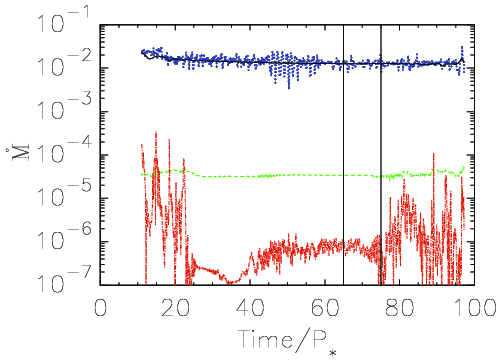}
\caption{Mass fluxes in the code units
$\dot{M}_0=\rho_{\mathrm d0}\sqrt{GM_\star R_\star^3}$ in the
various flow components in the simulation b5, shown in
Fig.~\ref{fig:sols1}. With vertical solid lines is
indicated the time interval in which we average the
fluxes in each of the flow components. With the solid
(black) line is shown the mass flux through the disk
at R=12R$_\star$ and with the dotted (blue) line the mass flux
loaded onto the star through the accretion column. Those two
fluxes are much larger than the fluxes in the other components
of the flow. The mass flux flowing through the magnetospheric
ejection at the radius R=12R$_\star$ is shown with the
dot-dashed (red) line, and the mass flux into the stellar
 wind from the vicinity of the stellar surface
is shown with the long-dashed (green) line.}
\label{fig:sols2}
\end{figure}
\begin{figure}
\includegraphics[width=\columnwidth]{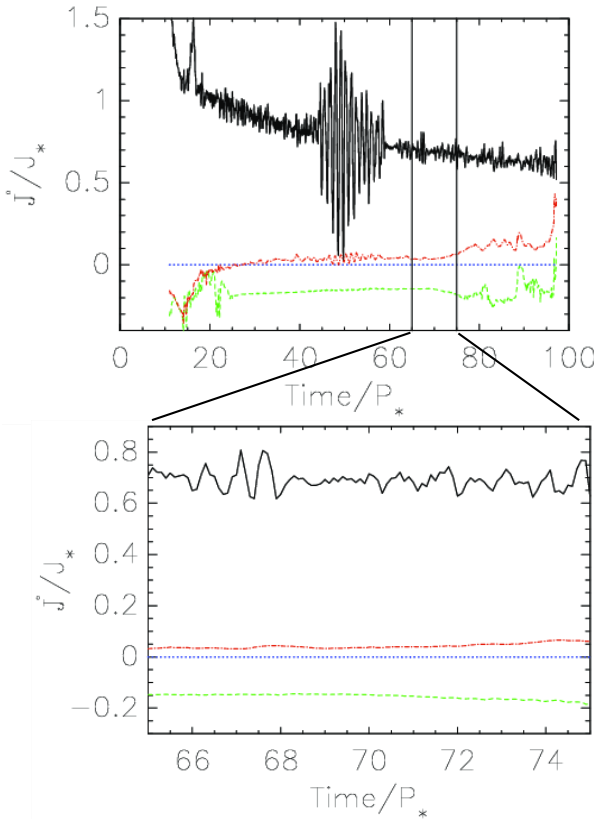}
\caption{Torques on the star, mostly exerted by the Maxwell stresses, in
the simulation b5, which is a DCE1 from Fig.~\ref{fig:sols1} with the
magnetospheric ejection, and mass fluxes shown in  \ref{fig:sols2}.
The quasi-stationary interval between the vertical lines is shown in detail
in the bottom panel. With the dashed (green) line is shown the torque by
the stellar wind. The torques by the matter flowing onto
the star through the accretion column from the distance beyond and below the
corotation radius $R_{\mathrm cor}$ are shown with the dotted
(blue) and solid (black) lines. With the dot-dashed (red) line is
shown the torque exerted on the star by the magnetospheric
ejection. Positive
torque spins the star up, and negative slows down its rotation. In this
case, the stellar rotation rate increases, so the star is spun up because
of the star-disk magnetospheric interaction. In the employed units of
$J_\star=M_\star R_\star^2 \Omega_\star$ the values correspond, in the case of
YSOs, to the stellar spin-up or spin-down in Myrs.}
\label{fig:sols2b}
\end{figure}
\begin{figure}
\includegraphics[width=0.9\columnwidth]{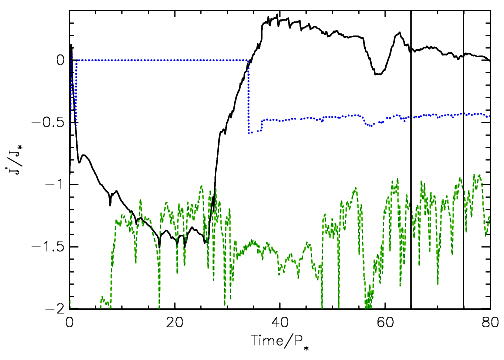}
\caption{Torques on the star in the simulation b16 (which is the DC case), with the same strength of magnetic field, 500~G as in the simulations b5 and b8, but with a star rotating two times faster. In this case, stellar rotation will be slowed down by the torque, because more of the torque comes from the
disk beyond the corotation radius $R_{\mathrm cor}$. The meanings behind the lines are the same as in the previous figure.
}
\label{slowdown}
\end{figure}
\begin{table}
\caption{We performed 64 star-disk magnetospheric interaction simulations
in a setup detailed in \cite{cem19}. There are all together 64 runs with
all the combinations of parameters as listed in the table. The magnetic
Prandtl number
P$_{\mathrm m}=\frac{2}{3}\alpha_{\mathrm v}/\alpha_{\mathrm m}$
is also listed -- in all the cases the anomalous viscosity parameter is
$\alpha_{\rm v}=1$. The four simulations shown in Fig.~\ref{fig:sols1}
are highlighted with boxed letters. Simulations in which
$\dot{J}_{\mathrm tot}>0$ are marked in bold. 
Annotated type of solution for each combination of parameters are shown in brackets, as
illustrated in Fig.~\ref{fig:sols1}. }
\centering                          % used for centering table
\begin{tabular}{ c c c c c }        % centered columns
\hline    % inserts single horizontal line
$\alpha_{\rm m}=$ & 0.1 & 0.4 & 0.7 & 1 \\
\hline\hline
$P_{\rm m}=$ & 6.7 & 1.67 & 0.95 & 0.67 \\
\hline\hline
 $\Omega_\star/\Omega_{\rm br}$ & & & & \\
\hline\hline
\multicolumn{5}{c}{$B_\star$=250~G} \\
0.05 & {\bf{a1}}(DCE1) & {\bf{a2}}(DC) & {\bf{a3}}(DC) & {\bf{a4}}(DC) \\
0.1 & {\bf{a5}}(DCE1) & {\bf{a6}}(DC) & a7(DC) & {\bf{a8}}(DC) \\
0.15 & {\bf{a9}}(DCE1) & {\bf{a10}}(DC) & a11(DC) & {\bf{a12}}(DC) \\
0.2 & a13(DCE1) & a14(DC) & a15(DC) & a16(DC) \\
\hline\hline     
\multicolumn{5}{c}{$B_\star$=500~G} \\
0.05 & {\bf{b1}}(DCE1) & {\bf{b2}}(DC) & {\bf{b3}}(DC) & {\bf{b4}}(DC) \\
0.1 & \boxed{\bf{b5}}(DCE1) & b6(DC) & b7(DC) & \boxed{\rm b8}(DC) \\
0.15 & {\bf{b9}}(DCE1) & b10(DC) & b11(DC) & b12(DC) \\
0.2 & b13(DCE1) & b14(DC) & b15(DC) & b16(DC) \\
\hline\hline    
\multicolumn{5}{c}{$B_\star$=750~G} \\
0.05 & {\bf{c1}}(DCE1) & c2(DC) & {\bf{c3}}(DC) & c4(DC) \\
0.1 & {\bf{c5}}(DCE1) & c6(DC) & c7(DC) & c8(DC) \\
0.15 & \boxed{\rm{c9}}(DCE2) & c10(DC) & c11(DC) & c12(DC) \\
0.2 & c13(DCE2) & c14(DC) & c15(DC) & c16(DC) \\
\hline\hline    
\multicolumn{5}{c}{$B_\star$=1000~G} \\
0.05 & {\bf{d1}}(DCE1) & d2(DC) & {\bf{d3}}(DC) & d4(DC) \\
0.1 & {\bf{d5}}(DCE1) & d6(DC) & d7(DC) & d8(DC) \\
0.15 & d9(DCE2) & d10(D) & d11(D) & \boxed{\rm d12}(D) \\
0.2 & d13(DCE2) & d14(D) & d15(D) & d16(D) \\
\hline\hline    
\end{tabular}
\label{params}
\end{table}
\begin{figure*}
\includegraphics[height=0.49\columnwidth,width=0.7\columnwidth]{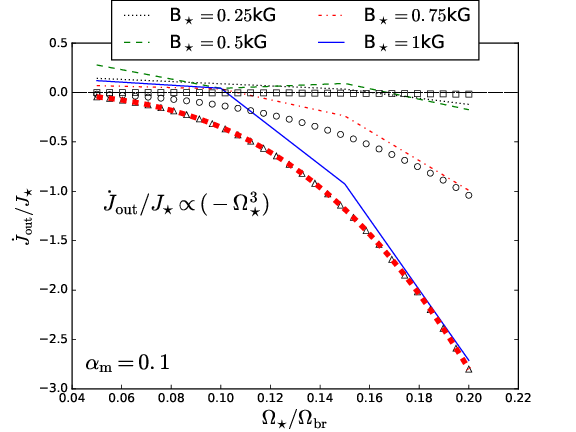}
\includegraphics[height=0.49\columnwidth,width=0.7\columnwidth]{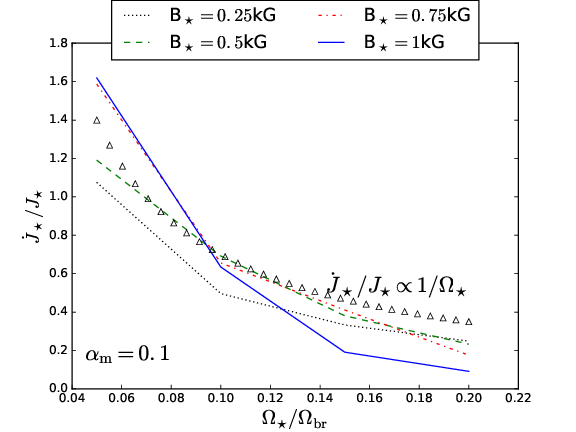}
\includegraphics[height=0.49\columnwidth,width=0.7\columnwidth]{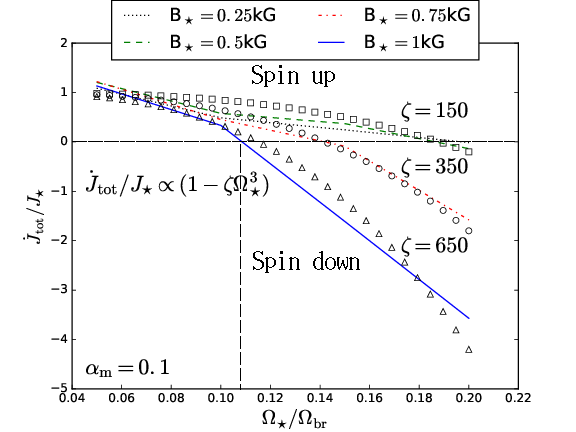}
\includegraphics[height=0.49\columnwidth,width=0.7\columnwidth]{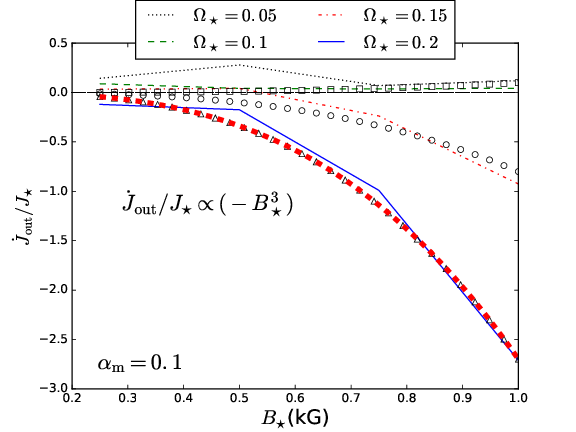}
\includegraphics[height=0.49\columnwidth,width=0.7\columnwidth]{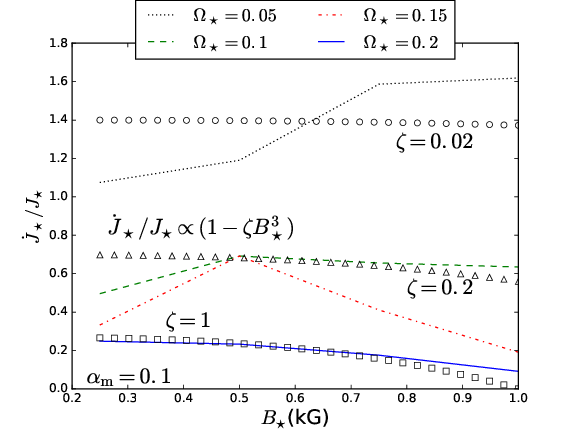}
\includegraphics[height=0.49\columnwidth,width=0.7\columnwidth]{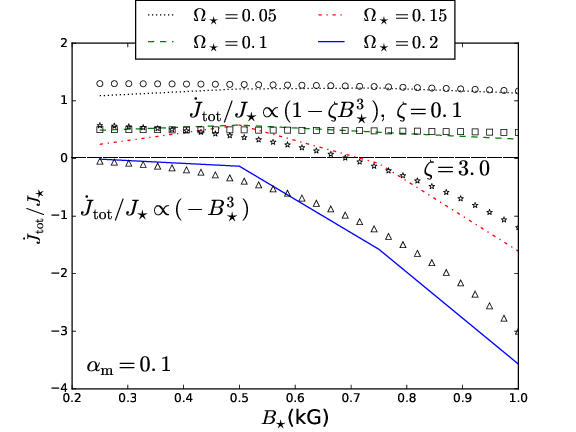}
\caption{Components of the torque (expressed in units of $J_\star$) exerted on the star in the cases a,b,c,d(1,5,9,13), with $\alpha_{\mathrm m}=0.1$, when a magnetospheric ejection is launched from the star-disk magnetosphere. With the square, circle and triangle symbols are plotted approximate matching
functions, indicated in each panel. Positive torque speeds the star up,
negative slows it down. {\it Top panels:} Trends in torque components for cases with different stellar magnetic field strengths, as a function of
the stellar rotation rate. {\it Bottom panels:} Same results
as in the top panels, but given as a function of stellar magnetic field
strength, organized by the different stellar rotation rates. Red lines in the left panels mark the examples of fitted lines (see text).
}
\label{jmes}
\end{figure*}
Similar simulations were previously reported in \cite{R09} with their
(not publicly available) code and in \cite{ZF09,ZF13} with the
(publicly available) PLUTO code (v.3) by \cite{m07}. Our simulations
in \cite{cpk17} were the first to repeat the latter setup, with the
updated version of the PLUTO code (v.4.1) and minor
amendments, as detailed in the appendix in \cite{cem19}. We used the same
resolution and choice of parameters to make sure we can make a direct comparison
in the parameter study with the results obtained in the previous
publications by \cite{ZF09,ZF13}.% and compiling the results.

We performed 64 star-disk magnetospheric interaction simulations
in a setup detailed in a spherical 2D axisymmetric
grid in a physical domain from the stellar surface to 30 stellar
radii. The resolution is at $(R\times\theta)=(217\times 100)$ grid cells,
with a logarithmic radial and uniform meridional distribution of grid
cells. We performed simulations in a quadrant of $\theta\in\lbrack
0,\pi/2\rbrack$, with the assumption of the equatorial symmetry. The
initial disk was set by the KK00 solution, with a non-rotating corona in
hydrostatic equilibrium. The viscosity coefficient was always set to
$\alpha_{\mathrm v}=1$, to avoid solutions with a midplane backflow;
in the analytical solution given in KK00, backflow appears in the cases with viscous
$\alpha$ parameter smaller than a critical value of 0.685. Solutions with
a midplane backflow we presented separately in \cite{MishraR20a,MishraR20b,MishraR22}.

After a few tens of stellar rotations, the system relaxes from the
initial and boundary conditions, and reaches the quasi-stationary state.
Examples of our results, showing the disk and its corona with the
accretion column and the magnetospheric ejection (``conical wind''
in \cite{R09} or ``conical outflow'' in \cite{cem19}) are shown in Fig.~\ref{fig:sols1}.

In our simulations, we systematically explored the parameter space shown
in Table~\ref{params}, with slowly rotating star, up to 20\% of the
stellar breakup rotation rate
$\Omega_{\mathrm br}=\sqrt{GM_\star/R_\star^3}$. In the case of YSOs, the probed stellar rotation periods were in
the range of 2-9 days, with the corresponding corotation radii
$R_{\mathrm cor}=(GM_\star/\Omega_\star^2)^{1/3}\sim3-7$
stellar radii (see Table~1 in \cite{cem19}). The second parameter we
varied was the stellar dipole magnetic field strength at the stellar
equator, $B_\star$, which can take values from 250-1000 Gauss\footnote{The
reference values we give as usually defined in the simulations community
and in the way they were given in the cited papers, instead referring to
the values nearby the pole, as usual in the observational community (we thank
to the anonymous referee for this notice).}. This makes our choice of field
strengths twice larger then expected fields observed in the YSO cases.
This does not change our conclusions: in our simulations shift to smaller
strengths of magnetic field preserves the trends well, numerical problems
arise with larger fields. The third varying parameter was the anomalous
resistivity coefficient, $\alpha_{\rm m}$, in the disk, which we set to
values ranging from 0.1 to 1.

With regard to the geometry of the solution and position of lines across
which we perform the integration of angular momentum and mass fluxes, the obtained solutions can be divided into the three cases as shown in
Fig.~2 in \cite{cem19}. For the reasons of additional analysis, here we
distinguish the results with magnetospheric ejections by the position of
the radius at which the ejection is launched: below or beyond the corotation
radius, resulting in four different states, as shown in Fig.~\ref{fig:sols1}.

 We exemplify the four distinct types of solutions with the representative
 simulations. In the simulation b8 shown in the top panel of Fig.~\ref{fig:sols1} we obtain a disk
 and accretion column. With a faster rotating star and larger magnetic field
 in the simulation d12, the disk is pushed away from the star and there is
 no accretion column-it is a disk only solution, a ``propeller'' regime
 \citep{Illarionov75,LovProp99}. These two cases echo a cartoon in
\cite{Matt05} (Fig.~3 in that work), which we also find in simulations. In the
two bottom panels are shown results with $\alpha_{\rm m}=0.1$, where in
addition to the disk and accretion column, we obtain magnetospheric
ejection: in the simulation b5 it is launched from below $R_{\mathrm cor}$,
and in simulation c9 beyond $R_{\mathrm cor}$.

Each of our 64 results can be presented as one of the four cases described
above, as shown in Fig.~\ref{fig:sols1}:
\begin{itemize}
\item[ ]{\bf DC:} (Disk+column) The disk inner radius and accretion column
are both positioned below the corotation radius. Stellar magnetic field lines are
anchored well beyond the corotation radius, $R_{\mathrm out}>R_{\mathrm cor}$.
\item[ ]{\bf D:} (Disk) With faster rotating star and larger magnetic field, the disk
inner radius is pushed further away from the star, beyond the corotation
radius, and the accretion column is not formed. This is the ``propeller
regime.'' The field lines are anchored in the disk far away from the star,
still enabling some inflow of matter onto the star, $R_{\mathrm out}>R_{\mathrm cor}$.
\item[ ]{\bf DCE1:} (Disk+Column+Ejection1) A disk truncation radius and
accretion column are both positioned below the corotation radius, and a
magnetospheric ejection is launched from the
magnetosphere. Stellar field is not reaching beyond the corotation radius,
$R_{\mathrm out}<R_{\mathrm cor}$.
\item[ ]{\bf DCE2:} (Disk+Column+Ejection2) The second type of solution with
a magnetospheric ejection, with the
disk truncation radius and accretion column positioned partly below, and partly
beyond the corotation radius. In this case, stellar field is anchored beyond
the corotation radius, $R_{\mathrm out}>R_{\mathrm cor}$, but just beyond
the accretion column footpoint in the disk.
\end{itemize}

The results with the different parameters are given in Table~2 in
\cite{cem19}. With the new distinction, in cases with magnetospheric
ejection, this table remains valid here; it is only the DCE cases that are split into DCE1
and DCE2 for the launching of the magnetospheric ejection below and beyond
the corotation radius, respectively. We list the solutions in Table~\ref{params}.

\section{Reach of the stellar magnetic field in the disk}\label{reach}
\begin{figure*}
\includegraphics[height=0.49\columnwidth,width=0.7\columnwidth]{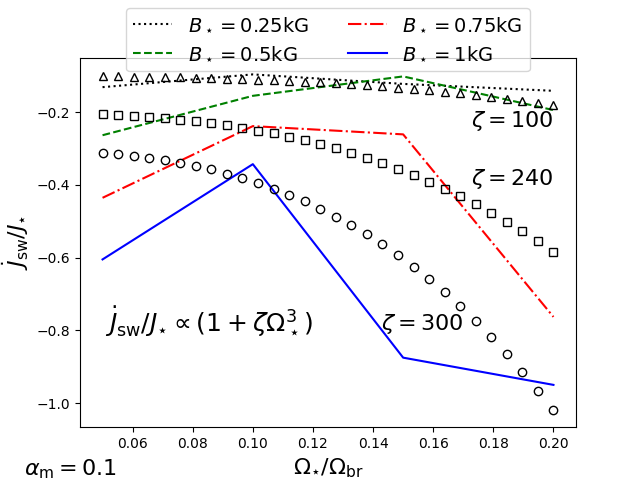}
\includegraphics[height=0.49\columnwidth,width=0.7\columnwidth]{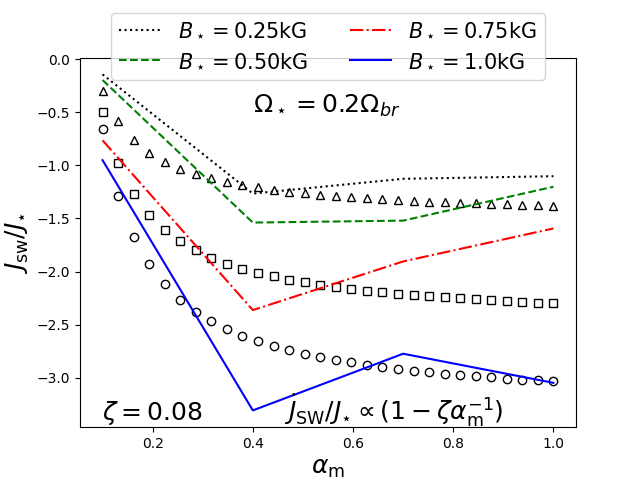}
\includegraphics[height=0.49\columnwidth,width=0.7\columnwidth]{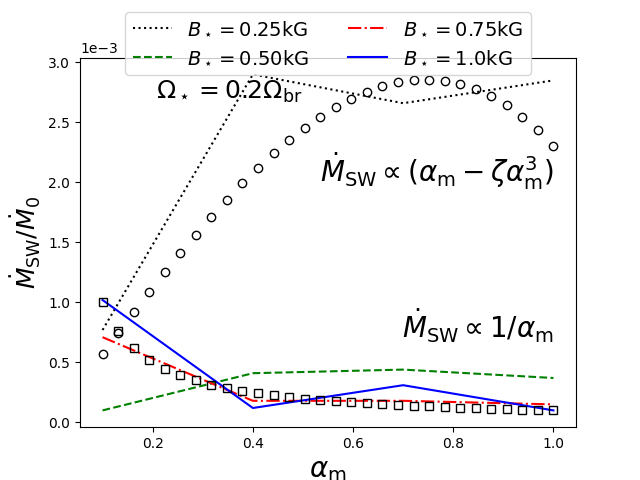}
\caption{Torques by the stellar wind, $\dot{J}_{\mathrm SW}$ (expressed in units of $J_\star$)
in terms of the dependence of the stellar rotation rate in the cases with the magnetospheric
ejection, $\alpha_{\mathrm m}=0.1$, ({\it left}). In all the cases except for the slowest stellar
rotation rates, $\dot{J}_{\mathrm SW}/J_\star$ drops with $\Omega_\star^3$ dependence.
$\dot{J}_{\mathrm SW}/J_\star$ is increasingly negative with the increase in stellar magnetic
field strength and rotation rate, and is also decreasing slightly with $1/\alpha_{\mathrm m}$ ({\it
middle}). Mass fluxes in the stellar wind in the cases with the fastest stellar rotation rates in
our simulations ({\it right}) also mostly decrease slightly with $1/\alpha_{\mathrm
m}$. The y-label multiplication factor is given in the left upper corner of this panel. }
\label{jswsa}
\end{figure*}
The characteristic radii which we can determine from our simulations
are the inner disk radius, $R_{\mathrm in}$, the corotation radius of the
material in the disk with the stellar surface (at the equator), $R_{\mathrm
cor}$, and the anchoring radius of the furthest line of magnetic field
connecting the disk with the stellar surface, $R_{\mathrm out}$.

Different characteristic radii have been discussed in the star-disk interaction
models, as, for instance, \cite{Matt05}. In Fig.~\ref{routrad}, we present our
results for the position of $R_{\mathrm out}$, where some trends
can be recovered: $R_{\mathrm out}$ increases with the larger resistive
coefficient $\alpha_{\mathrm m}$ for all stellar magnetic fields. There
are some significant departures from the trends, which are probably
related to the details of the flow geometry: in the case with
$B_\star=0.5~kG$ (shown in the right top panel in the same figure) the
largest $R_{\mathrm out}$ is measured at the slowest stellar rotation
rate, out of the trend for other magnetic field strengths. In the
bottom panels in this figure are shown the inner disk radii, $R_{\mathrm in}$,
with the same parameters, showing similar trends with the resistivity coefficient,
$\alpha_{\mathrm m}$, and stellar magnetic field.

In all the cases with $\alpha_{\mathrm m}=0.1$, a magnetospheric
ejection is launched from the system in our simulations. In Fig.~\ref{routradb}, it
is shown that in such cases, there is only a minor dependence of
$R_{\mathrm out}$ on the stellar rotation rates for all the stellar
field strengths. Also, the increase of $R_{\mathrm out}$ with the stellar
field strength is not large in such cases. We thus go on to consider why the magnetospheric
ejections are launched only in cases where $\alpha_{\mathrm m}=0.1$? With larger values of $\alpha_{\mathrm m}$, there is obviously enough magnetic
diffusion to allow the matter to cross the magnetic field lines not to push
them towards the star during accretion. When there is not enough dissipation,
as with $\alpha_{\mathrm m}=0.1$, the magnetic field lines will be pressed
towards the star, where the mounting magnetic pressure pushes them away from
the star, expelling some of matter in the magnetospheric ejections.
The DCE2 type of solution from Fig.~\ref{fig:sols1} (Sim. c9), pushed further
in the magnetic field strength and with more magnetic diffusivity, would
finish in the ``propeller regime,'' type D (Sim. d12).
 
\section{Torque exerted on a star}\label{angmom}
\begin{figure*}
\includegraphics[height=0.5\columnwidth,width=0.7\columnwidth]{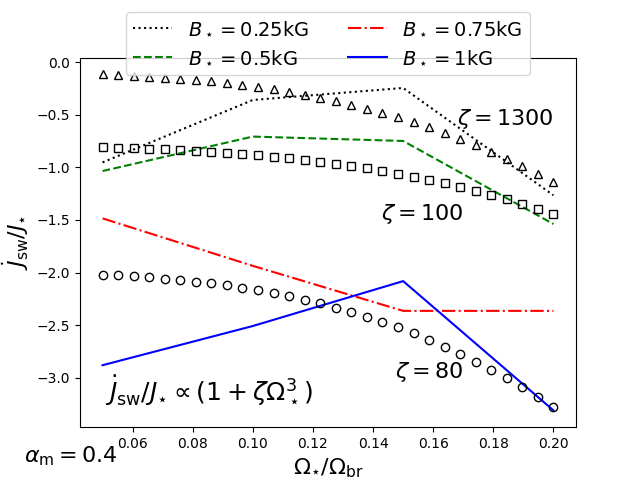}
\includegraphics[height=0.5\columnwidth,width=0.7\columnwidth]{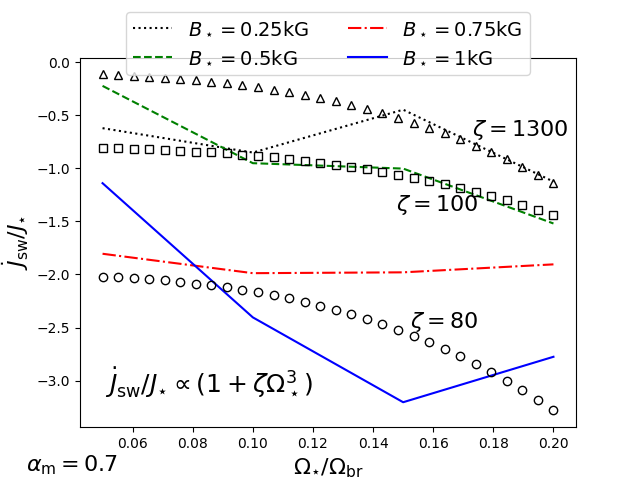}
\includegraphics[height=0.5\columnwidth,width=0.7\columnwidth]{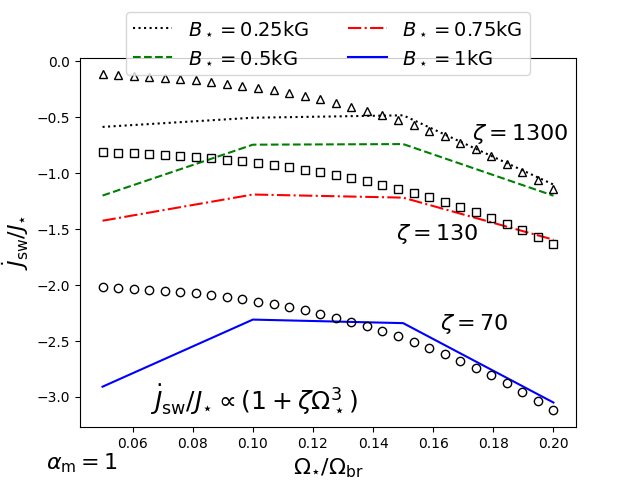}
\caption{Torques by the stellar wind $\dot{J}_{\mathrm SW}$ (expressed in units of $J_\star$) in dependence of the stellar rotation rate in the cases without magnetospheric
ejection ($\alpha_{\mathrm m}=0.4, 0.7, 1.0$). With the square, circle and triangle symbols are plotted approximate matching functions, indicated in each panel. In most of the cases, $\dot{J}_{\mathrm SW}$ is increasingly negative with the increase in stellar magnetic field strength and rotation rate. The cases with slower stellar rotation  often do not match the approximated functions.}
\label{jsws}
\end{figure*}
\begin{figure*}
\includegraphics[height=0.5\columnwidth,width=0.7\columnwidth]{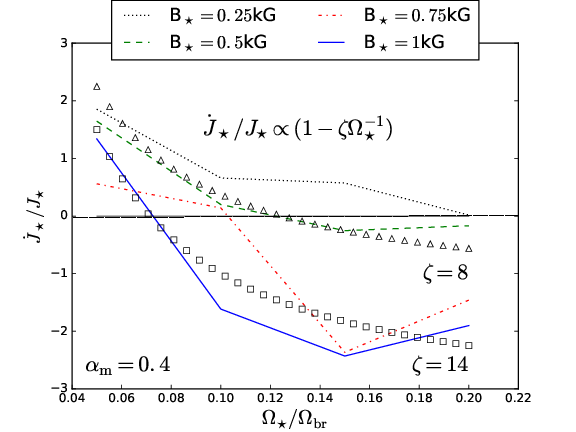}
\includegraphics[height=0.5\columnwidth,width=0.7\columnwidth]{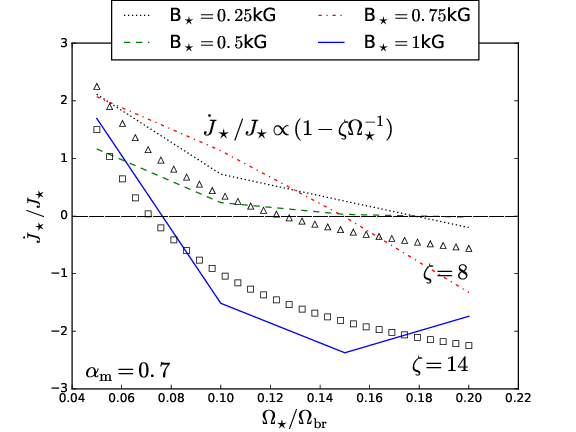}
\includegraphics[height=0.5\columnwidth,width=0.7\columnwidth]{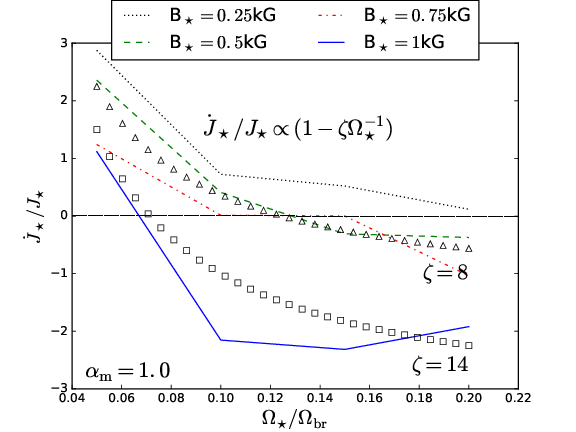}
\includegraphics[height=0.5\columnwidth,width=0.7\columnwidth]{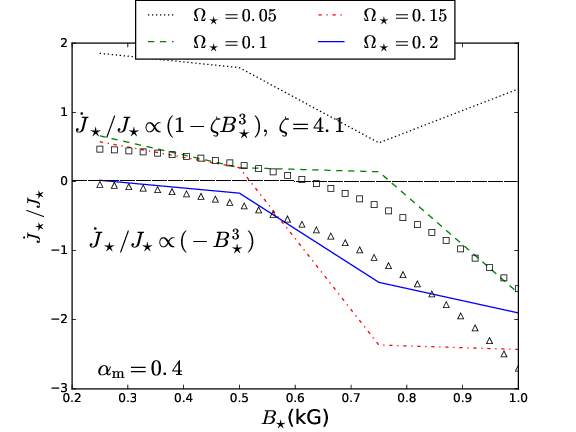}
\includegraphics[height=0.5\columnwidth,width=0.7\columnwidth]{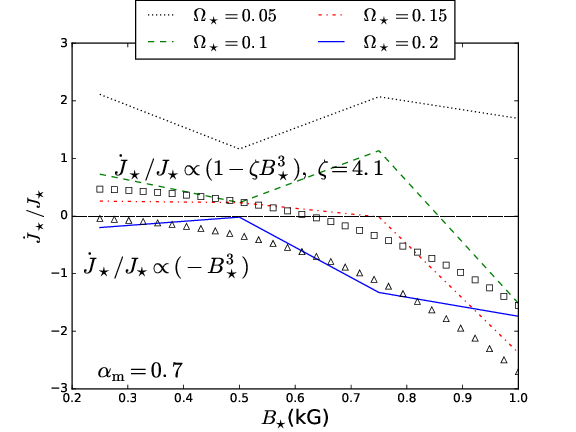}
\includegraphics[height=0.5\columnwidth,width=0.7\columnwidth]{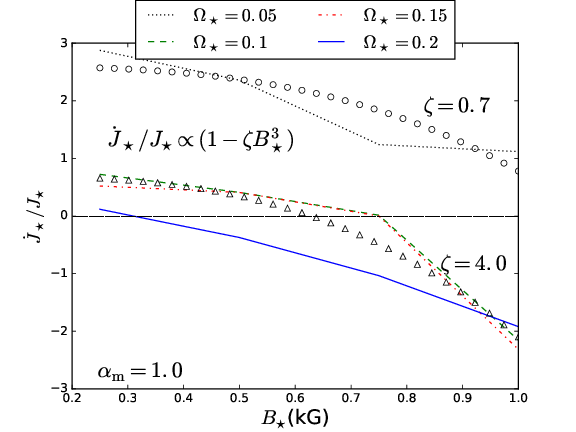}
\caption{Torques, $\dot{J}_\star$, (expressed in units of $J_\star$) exerted on the star by material infalling from the disk in all the cases without magnetospheric ejection:\ these are all the cases except
a,b,c,d(1,5,9,13), with the resistive coefficient $\alpha_{\mathrm m}$=0.4, 0.7 and 1.0. Approximate matching functions and trends in solutions with different stellar rotation rates and magnetic
field strengths are shown as $\dot{J}_\star(\Omega_\star)/J_\star$ ({\it top})
and $\dot{J}_\star(B_\star)/J_\star$ ({\it bottom}). The dependence on
$\alpha_{\mathrm m}$ is small.
}
\label{jstar2}
\end{figure*}
\begin{figure*}
\includegraphics[height=0.5\columnwidth,width=0.7\columnwidth]{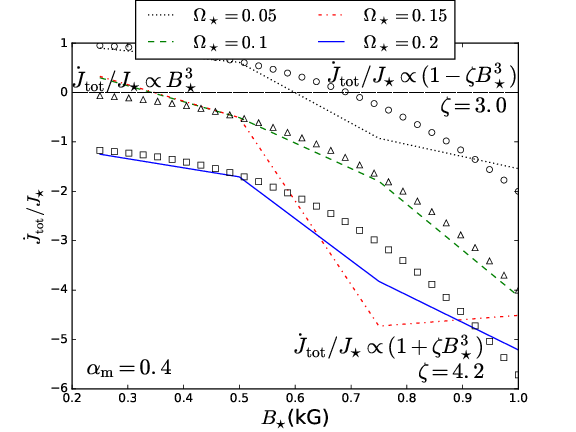}
\includegraphics[height=0.5\columnwidth,width=0.7\columnwidth]{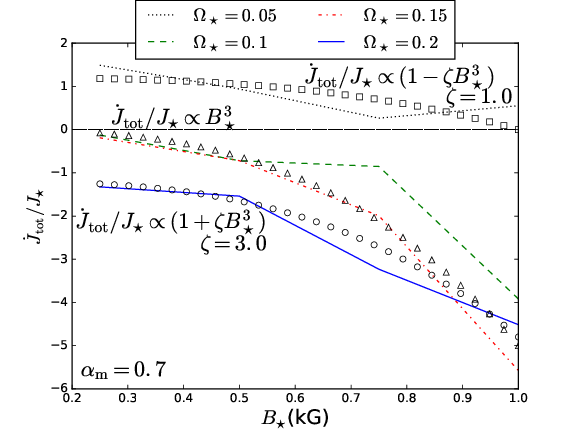}
\includegraphics[height=0.5\columnwidth,width=0.7\columnwidth]{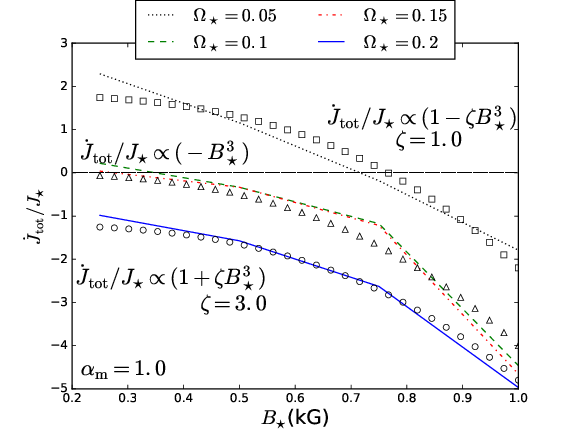}
\caption{Total torques in the system $\dot{J}_{\mathrm tot}(B_\star)$,
(expressed in units of $J_\star$) in the cases without magnetospheric
ejection from
Fig.~\ref{jstar2}--all but a,b,c,d(1,5,9,13). Results with the same
resistive coefficient, $\alpha_{\mathrm m}$, are shown together and
corresponding matching functions are outlined. The dependence on
$\alpha_{\mathrm m}$ is small. }
\label{jtot}
\end{figure*}
\begin{figure*}
\includegraphics[height=0.5\columnwidth,width=0.7\columnwidth]{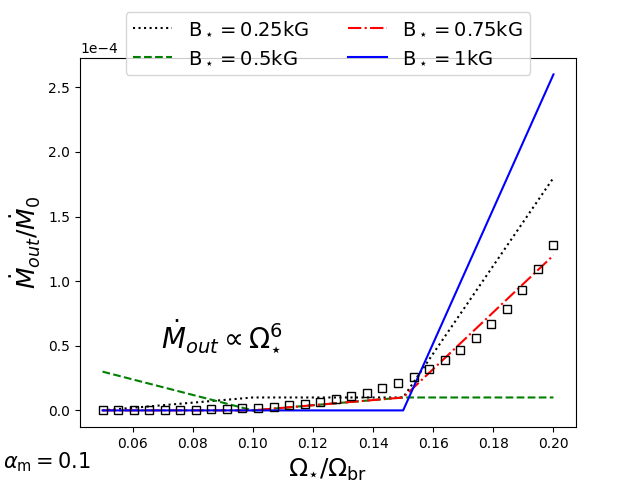}
\includegraphics[height=0.5\columnwidth,width=0.7\columnwidth]{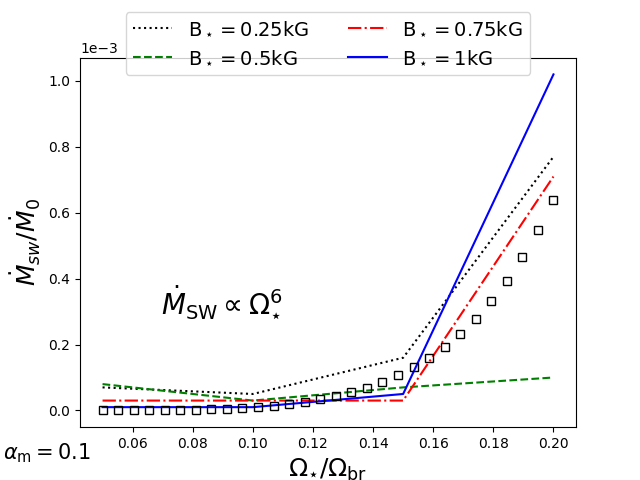}
\includegraphics[height=0.5\columnwidth,width=0.7\columnwidth]{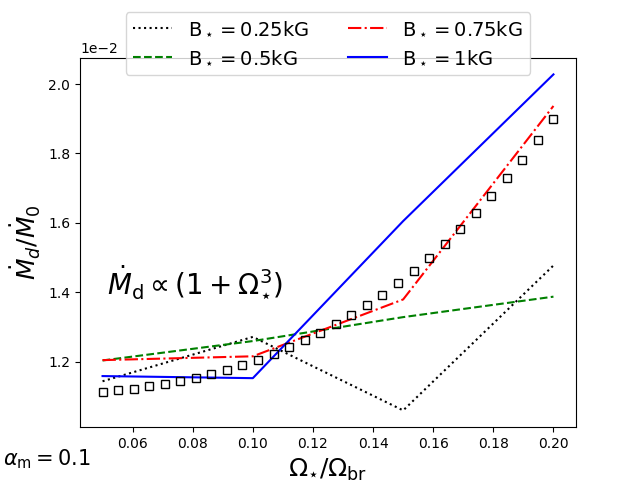}
\includegraphics[height=0.5\columnwidth,width=0.7\columnwidth]{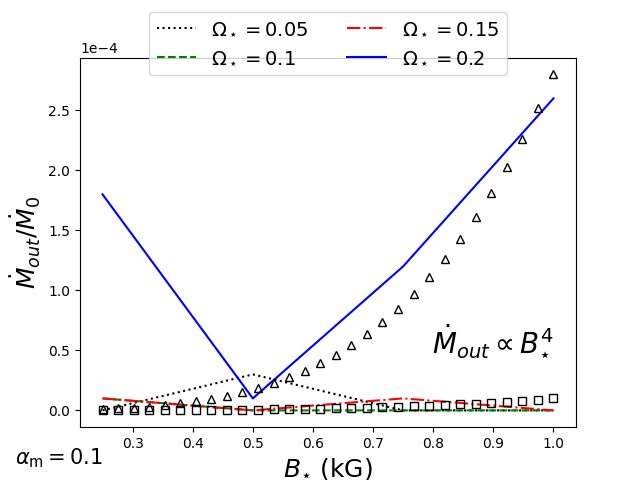}
\includegraphics[height=0.5\columnwidth,width=0.7\columnwidth]{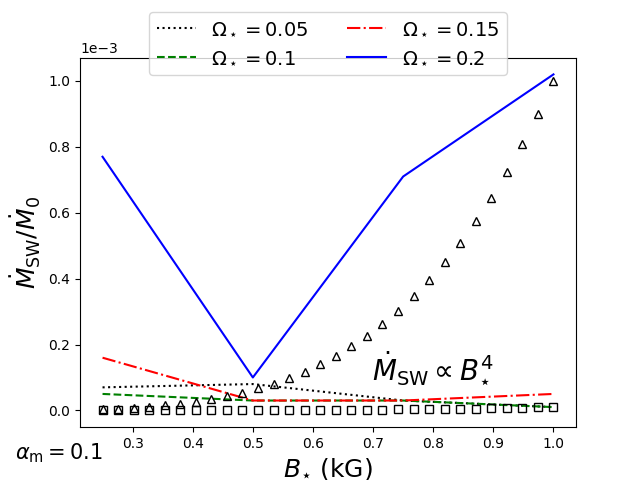}
\includegraphics[height=0.5\columnwidth,width=0.7\columnwidth]{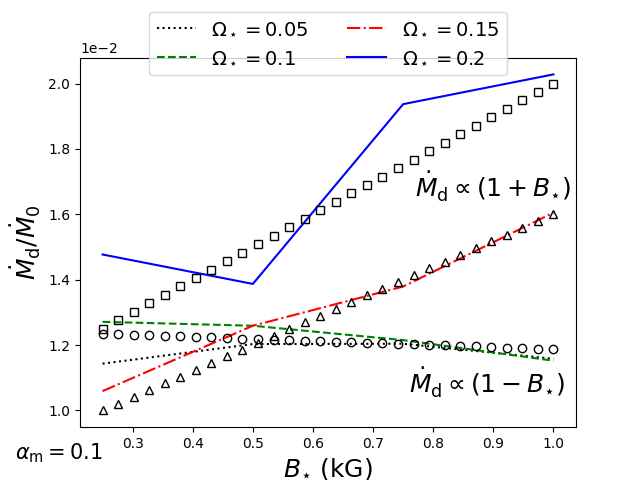}
\includegraphics[height=0.5\columnwidth,width=0.7\columnwidth]{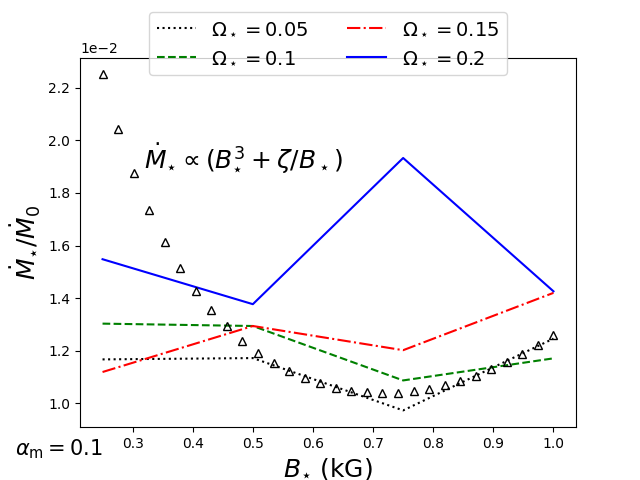}
\includegraphics[height=0.5\columnwidth,width=0.7\columnwidth]{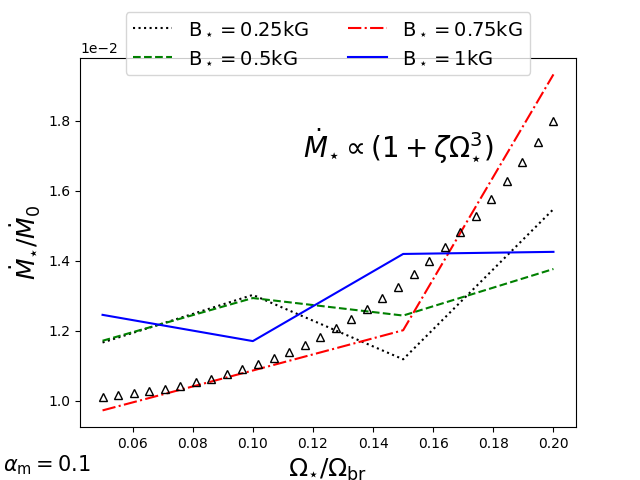}
\caption{Mass losses in the outflow and in the stellar wind, $\dot{M}_{\mathrm out}$ and $\dot{M}_{\mathrm SW}$, and mass flux through the disk $\dot{M}_{\mathrm d}$ in solutions with magnetospheric outflow ($\alpha_{\mathrm m}=0.1$), in terms of the dependence of the stellar rotation rate ({\it top}) and of the strength of stellar magnetic field ({\it middle}). The mass flux onto the star $\dot{M}_{\star}$ is also shown{\it (bottom)}. The values of $\dot{M}_{\mathrm out}$ and $\dot{M}_{\mathrm d}$ are computed at R=12$R_\star$, while $\dot{M}_{\mathrm SW}$ and $\dot{M}_{\star}$ are computed at the stellar surface. For illustration, shown are examples of matching functions. Note: the y-label multiplication factor is given in the left upper corner of the panels.
}
\label{mdotoutsfits}
\end{figure*}
\begin{figure*}
\includegraphics[height=0.5\columnwidth,width=0.7\columnwidth]{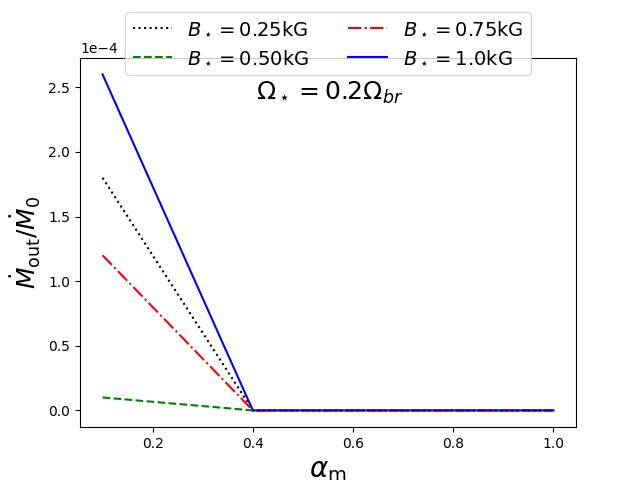}
\includegraphics[height=0.5\columnwidth,width=0.7\columnwidth]{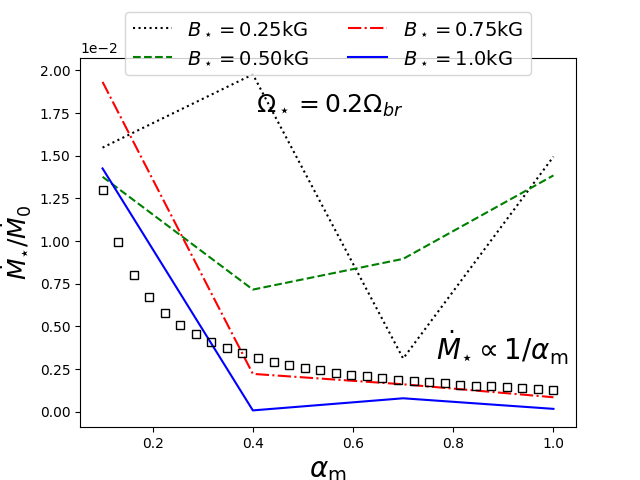}
\includegraphics[height=0.5\columnwidth,width=0.7\columnwidth]{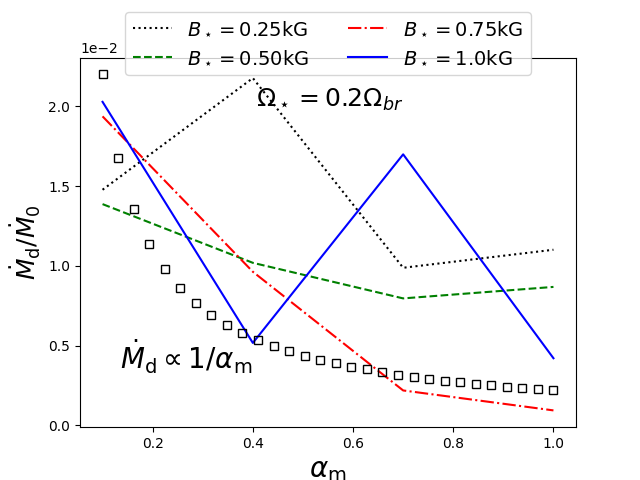}
\caption{Components of the mass fluxes, $\dot{M,}$ in terms of the dependence of the anomalous resistive coefficient, $\alpha_{\mathrm m}$, for different strengths of the stellar magnetic field in the cases with fastest stellar rotation in our simulations, $\Omega_\star=0.2\Omega_{\mathrm br}$. The values of $\dot{M}_{\mathrm out}$ and $\dot{M}_{\mathrm d}$ are both computed at R=12$R_\star$, where the flow is most stable. The latter is computed across the disk height, and corresponds to the total mass accretion rate available for distribution in the system. The component $\dot{M}_{\mathrm{SW}}$ at the same stellar rotation rate is shown together with corresponding torques in Fig.~\ref{jswsa}. Note: the y-label multiplication factor is given in the left upper corner of the panels. }
\label{mdotsom020}
\end{figure*}
\begin{figure}
\includegraphics[width=0.9\columnwidth]{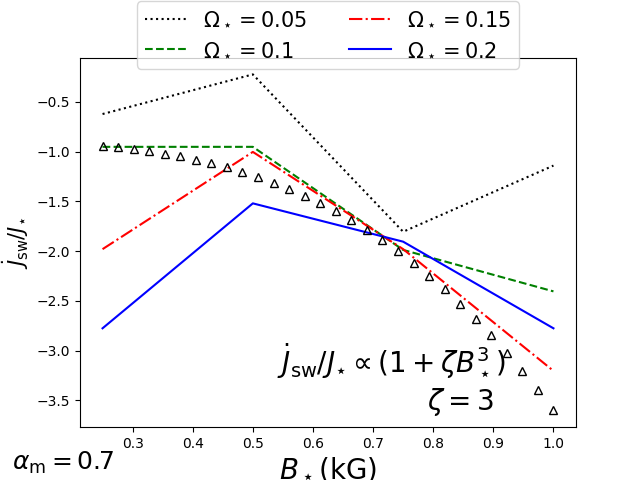}
\caption{Torques by the stellar wind $\dot{J}_{\mathrm sw}(B_\star)$
(expressed in units of $J_\star$) in cases a,b,c,d(3,7,11,15) with
$\alpha_{\mathrm m}=0.7$. Triangle symbols represent
approximate matching function. The torque by the stellar wind is in most of those cases increasingly negative with increasing stellar
rotation rates.}
\label{jswofalpha}
\end{figure}
Next, we compute the torques applied on the star from the various
components of the flow and magnetic field. The azimuthal magnetic
field plays a key role in the torques computed, so in Fig.~\ref{bphis}
we illustrate the behavior of $B_\varphi$ with respect to the stellar
rotation rate, based on the example of solutions without a magnetospheric
ejection,
with $\alpha_{\mathrm m}$=1 and $B_\star=500$~G. We find that
$B_\varphi$ atop the disk (top panel) is small and does not vary much
in the cases with different $\Omega_\star$.

The mass and angular momentum fluxes are obtained by integrating
\begin{eqnarray}
\dot{M}=\int_{\rm S}\rho\vec{\rm v}_{\rm p}\cdot
d\vec{S},\quad\quad\quad\quad\\
\dot{J}_{\mathrm tot}=\int_{\rm S}\left(r\rho v_\varphi\vec{\rm v}_{\rm p}-\frac{rB_\varphi\vec{B}_{\rm p}}{4\pi}\right)d\vec{S},
\label{mjdot}
\end{eqnarray}
where $\vec{S}$ is the surface of integration.

We computed the fluxes in each of the 64 simulations, to find trends in
different combinations of parameters. Depending on the geometry of the
solution, the boundaries of the integration change, as shown in
Fig.~\ref{fig:sols1}. We discuss the possible configurations below.

First, the simplest configuration is with the disk and
accretion column, which we find in majority of the simulations with the
resistivity coefficient $\alpha_{\mathrm m}>0.1$. Then, in cases with
larger magnetic field and faster stellar rotation, we find that the
disk is pushed away from the star, and the accretion column is disrupted.
Finally, in cases with a magnetospheric ejection, there is an additional
component in the flow, so we then modify the contours of integration.

The mass accretion rates are computed across the various surfaces. The mass load in the stellar wind is:
\begin{equation}
\dot{M}_{\mathrm SW}=4\pi R_\star^2\int_{\theta_{\mathrm a}}^0\rho v_{\mathrm R}\sin\theta a\theta ,
\label{mdotsw}
\end{equation}
computed across the stellar surface. The accretion rate onto the stellar surface
\begin{equation}
\dot{M}_\star=-4\pi R_\star^2\int_{\theta_{\mathrm b}}^{\pi/2}\rho v_{\mathrm R}\sin\theta d\theta, 
\label{mdotstar}
\end{equation}
is also computed across the stellar surface. The $\theta_{\mathrm a}$ refers to the angle of the last
open magnetic surface and $\theta_{\mathrm b}$ to the footpoint of the last closed line of magnetic
flux  {\bf b} (see, e.g., the third panel in Fig.~\ref{fig:sols1}, for the DCE1
case). The disk accretion rate,
\begin{equation}
\dot{M}_{\mathrm d}=-4\pi R^2\int_{\theta_{\mathrm b}}^{\pi/2}\rho v_{\mathrm R}\sin\theta d\theta, 
\label{mdotdisk}
\end{equation}
is computed at R=12R$_\star$, the same as mass load in the magnetospheric ejection,
\begin{equation}
\dot{M}_{\mathrm out}=4\pi R^2\int^{\theta_{\mathrm a}}_{\theta_{\mathrm a'}}\rho v_{\mathrm R}\sin\theta d\theta ,
\label{mdotout}
\end{equation}
where $\theta_{\mathrm disk}$ is the angle at which is the disk surface, $\theta_{\mathrm a}$ and $\theta_{\mathrm a'}$ are the angles encompassing the magnetospheric ejection\footnote{Components in the magnetospheric ejection can be divided into part related to the star and the disk. Detailed account is given in \cite{ZF13}, here we do not use this distinction.}, as shown in  Fig.~\ref{fig:sols1}. The different distances at which the mass fluxes are computed for separate components in the flow render the sum for the total mass flux elusive, as some of the mass flux is mixed between the components between the stellar surface and R=12R$_\star$. The relative discrepancy in the mass accretion rate in the disk and the sum of the mass fluxes in our results, $(\dot{M}_{\mathrm d}-\dot{M}_{\mathrm SW}-\dot{M}_\star-\dot{M}_{\mathrm out})/\dot{M}_{\mathrm d}$, measures this mixing. In most of our cases the relative discrepancy is low, below 10\%. In the cases with violent reconnection when the flow is severely disrupted at some locations during the run, or with some other instability, it can grow, up to few tens of percents. This is probably the reason for some of the outliers in the trends for mass fluxes. In the subsections in Sect.~\ref{trends}, we show that mass fluxes enter the (approximate) analytical solutions for torques, through the coefficients of proportionality in the expressions for different characteristic radii. To better capture the different physical regimes in the flows, a separation of the results from our simulations in more sub-groups would probably be needed. We leave this aspect for a future study.

The torque from the stellar surface into the
stellar wind,\footnote{For shortness and to avoid confusion with the
magnetospheric ejections, we abbreviate SW, but this outflow is
not from the stellar surface, which is an absorbing boundary in our
simulations. It is from a material from the disk, diverted away from the
star by the magnetospheric interaction.} $\dot{J}_{\mathrm SW}$, is
computed above the line {\bf a}. Torque on the star, exerted by the disk,
is computed below the line {\bf b}, where
$\dot{J}_{\mathrm R>R_{\mathrm cor}}$ is part of the torque beyond the
corotation radius $R_{\mathrm cor}$, reaching to the line {\bf c},
below which the contribution is $\dot{J}_{\mathrm R<R_{\mathrm cor}}$.
Depending on the position of $R_{\mathrm cor}$, one or both of those
contributions to $\dot{J}_{\mathrm tot}$ are present.

In the cases with a magnetospheric ejection, $\dot{J}_{\mathrm out}$
is computed between the lines {\bf a} and {\bf b} (with the ejection
confined between the lines {\bf a} and {\bf a'}, between which the
reconnection occurs). Note our use of the $\dot{J}_{\mathrm out}$ (with {\it out} for {\it outflow}) for the magnetospheric ejections, for consistency with our previous publication \citep{cem19}, instead of $\dot{J}_{\mathrm ME}$, used by \cite{ZF13}. In the magnetospheric ejection, the part
beyond the $R_{\mathrm cor}$ also slows down the star, and the part
below $R_{\mathrm cor}$ spins the star up\footnote{As shown in
\cite{LN13}, details of the magnetic field inside the disk can
complicate this simple picture.}.

In \cite{ZF09,ZF13} is provided a detailed discussion of the mass, angular
momentum and energy fluxes in the simulations of star-disk magnetospheric
interaction. Here we focus on the torques in various flow components
in the system, with different physical parameters. Results from our
parameter study will help to find if there are trends in contributions
in the flows to the spinning up or down of the central object, with
respect to the parameters varied in the simulations.

The torques are computed by integrating the expression
$\Lambda=R_\star^3(-4\pi\rho v_Rv_\varphi+B_RB_\varphi)
\sin^2\theta$ along the different parts of the stellar surface\footnote{The
angles 0, $\theta_{a,b,c}$, and $\pi/2$ are taken along the circle at stellar
radius, $R_\star$.}. The first term in $\Lambda$ is the kinetic torque,
which is found to be negligible in all the cases. This leaves us with mostly
the second term, namely, magnetic Maxwell stresses, contributing to the stellar
magnetic torque. 

Integration is performed along the four segments of stellar surface:
\begin{eqnarray}
\dot{J}_{\mathrm SW}=\int_{0}^{\theta_{a}}\Lambda d\theta,\ 
\dot{J}_{\mathrm out}=\int_{\theta_{a}}^{\theta_{b}}\Lambda d\theta,\ \nonumber\\
\dot{J}_{\mathrm R>R_{\mathrm cor}}=\int_{\theta_{b}}^{\theta_{c}}\Lambda d\theta,\ 
\dot{J}_{\mathrm R<R_{\mathrm cor}}=\int_{\theta_{c}}^{\pi/2}\Lambda d\theta .
\label{jdots}
\end{eqnarray}
If we introduce $\dot{J}_\star=\dot{J}_{\mathrm R>R_{\mathrm cor}}+\dot{J}_{\mathrm R<R_{\mathrm
cor}}$, we can write the total torque as:
\begin{equation}
\dot{J}_{\mathrm tot}=\dot{J}_{\mathrm SW}+\dot{J}_{\mathrm out}+\dot{J}_\star .\quad\quad\quad\quad\quad
\label{jtoteq}
\end{equation}
Here, $\dot{J}_{\mathrm SW}$ is computed over the area threaded by
the opened field lines and $\dot{J}_{\mathrm out}$ over the magnetospheric
ejection.
$\dot{J}_{\mathrm R>R_{\mathrm cor}}$ accounts for the matter from the disk
which is originating beyond the corotation radius $R_{\mathrm cor}$, and
$\dot{J}_{\mathrm R<R_{\mathrm cor}}$ for the matter from the disk
originating below the $R_{\mathrm cor}$.

The sign convention is such that a positive angular momentum flux spins
the star up and a negative slows its rotation down. The whole meridional
plane is taken into account by multiplying the result by 2 for the
symmetry across the disk equatorial plane. We normalize the torque
to total stellar angular momentum $J_\star=I_\star\Omega_\star$, with the
stellar moment of inertia $I_\star=k^2M_\star R_\star^2$, where $k^2=0.2$
is the typical normalized gyration radius of a fully convective star. With
such a normalization, the inverse of the characteristic scale for change of
stellar rotation rate is readily obtained as proportional to
$B_\star^2M_\star^{-3/2}R_\star^{5/2}$. For the typical YSOs, the scale for
$\dot{J}_\star/J_\star$ in our plots approximately corresponds to Myrs.

Examples of the mass and angular momentum fluxes throughout our
simulation are given in Figs.~\ref{fig:sols2}-\ref{slowdown}. Mass fluxes
through the disk and onto the star are, in the YSO case, about
$5\times 10^{-9}M_\sun yr^{-1}$. An interval is marked with the vertical
solid lines, in which both the mass and angular momentum fluxes are not
varying much, as shown in Fig.~\ref{fig:sols2b}. We computed an average of
the angular momentum flux between those lines for the cases with various
parameters. Then we compared the values obtained in the different cases. In the
example case, the star is spun up.

A case when a stellar rotation is being slowed down is shown in
Fig.~\ref{slowdown}. Stellar magnetic field in this case is of the same
strength as in the previous example, but the star is rotating faster,
and it turns out that more of the torque on the stellar surface comes
from the region in the disk beyond the corotation radius.

\section{Trends in torques acting on a star}\label{trends}
\begin{figure*}
\includegraphics[height=0.5\columnwidth,width=0.7\columnwidth]{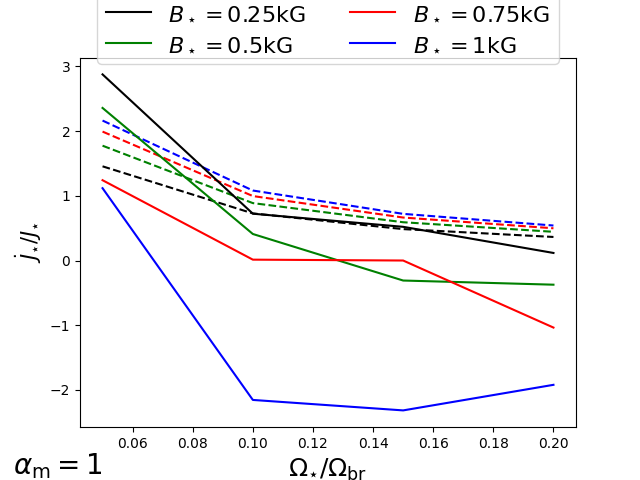}
\includegraphics[height=0.5\columnwidth,width=0.7\columnwidth]{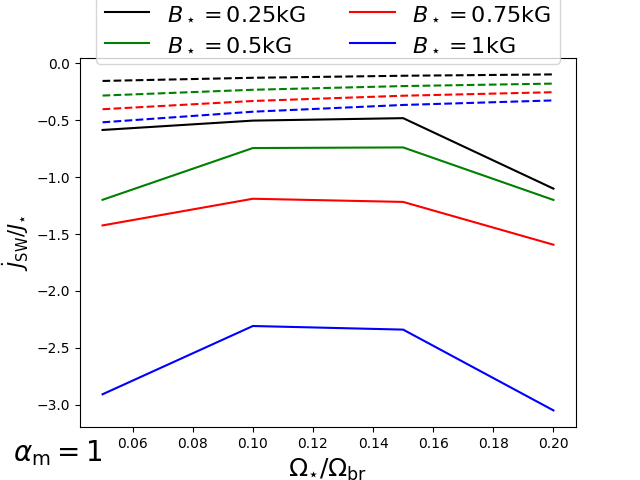}
\includegraphics[height=0.5\columnwidth,width=0.65\columnwidth]{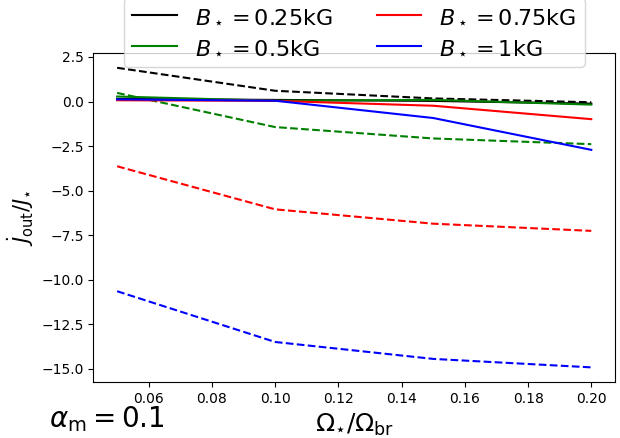}
\includegraphics[height=0.5\columnwidth,width=0.7\columnwidth]{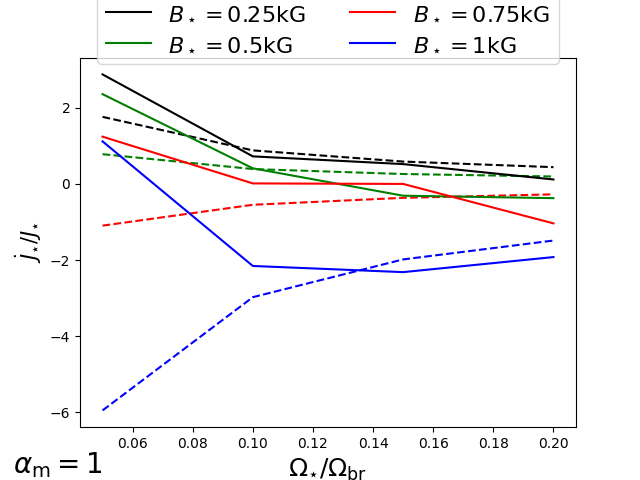}
\includegraphics[height=0.5\columnwidth,width=0.7\columnwidth]{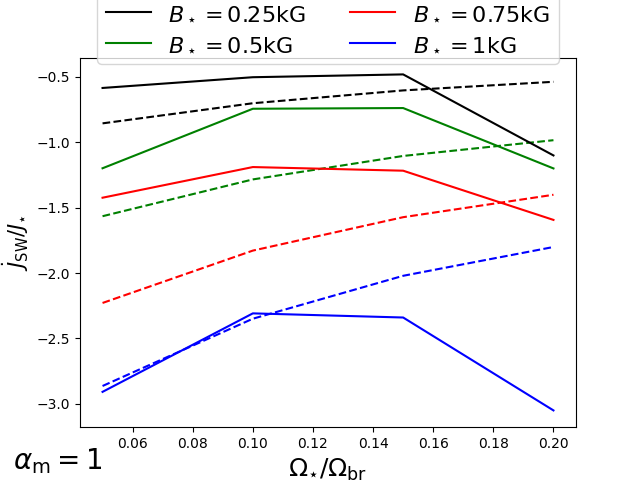}
\includegraphics[height=0.5\columnwidth,width=0.65\columnwidth]{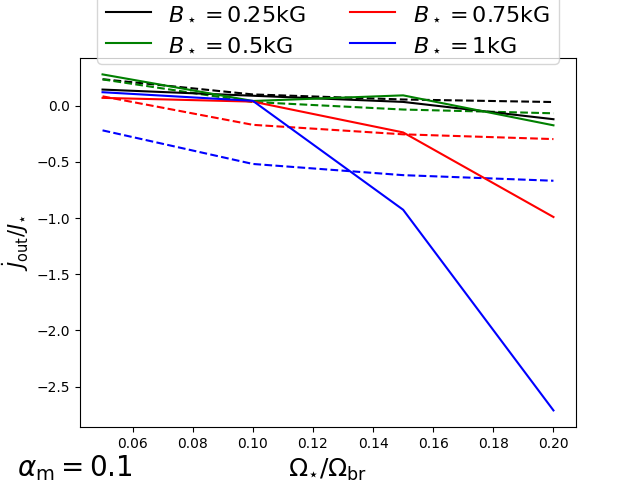}
\caption{Comparison of our results (shown in solid lines) with the results from
\cite{GalZanAma19} (shown in dashed lines) expressed in our units of $J_\star$ ({\it top}).
The line colors correspond to the same magnetic field strengths in both solid
and dashed lines. In the leftmost panel, $K_{\mathrm acc}=1$ was
set in Eq.~\ref{eq4GZA}, and $K_1=1.7$ and $K_2=0.0506$, $m=0.2177$ in Eqs.~\ref{ravalf} and \ref{eq4GB}. The results from our simulations differ from \cite{GalZanAma19} predictions, in which they assumed those factors to be constant. In our simulations, those quantities are self-consistently adjusting. Examples of curves ({\it bottom}) for the same models by \cite{GalZanAma19} as in the top panels (shown with the dashed lines) with the accretion factors $K_{\mathrm acc}$, $K_{\mathrm ME}$, and $K_1$ modified in such a way to (at least in some of the solutions) better match the results from our simulations, shown with the solid lines. The text gives details on the modifications.}
\label{jcomppGB19}
\end{figure*}
\begin{figure}
\includegraphics[width=0.9\columnwidth]{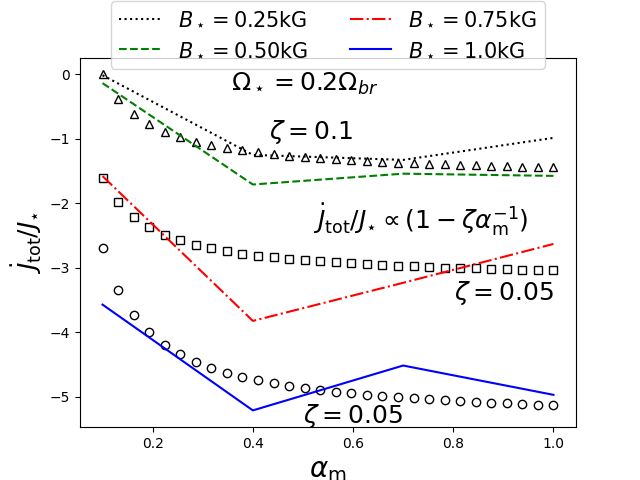}
\includegraphics[width=0.9\columnwidth]{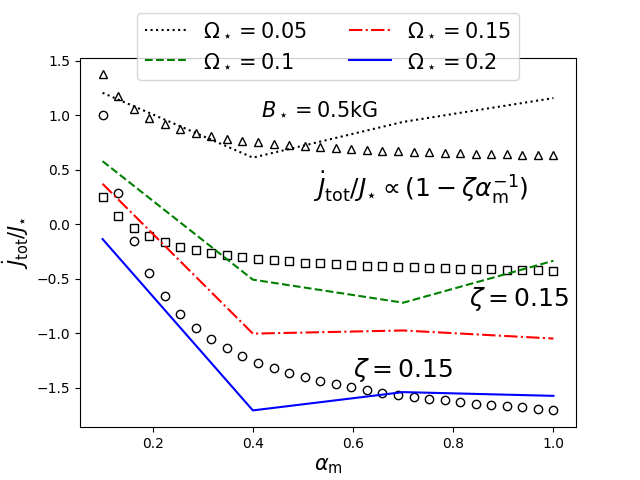}
\caption{Total torques in the system $\dot{J}_{\mathrm tot}(\alpha_{\mathrm m})$ 
(expressed in units of $J_\star$) in cases a,b,c,d(13,14,15,16),
with fastest stellar rotation in our sample, $\Omega_{\star}=0.2$ ({\it top}) and in cases b(1-16) with B$_\star=0.5$~kG ({\it bottom}), showing the $1/\alpha_{\mathrm m}$ dependence. Similar results are obtained in most of the other cases.}
\label{jtotomalph}
\end{figure}
\begin{figure}
\includegraphics[width=0.9\columnwidth]{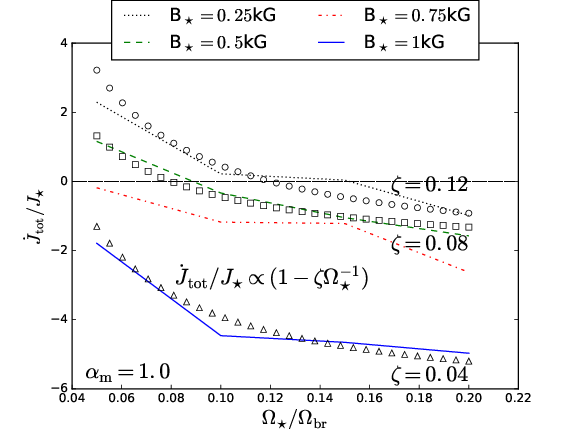}
\caption{Total torques (expressed in units of $J_\star$) for the anomalous
resistivity coefficient $\alpha_{\mathrm m}=1.0$ in the cases
a,b,c,d(4,8,12,16) with different $B_\star$, in terms of the dependence of
stellar rotation rate, $\dot{J}_{\mathrm tot}(\Omega_\star)$. The total
torque in the system is increasingly negative with the increase in
stellar field, $B_\star$.}
\label{jtotoom}
\end{figure}
We go on to analyze the results from our simulations to find trends in the
solutions with different parameters. The torques are computed in
the cases with varying strengths of the stellar magnetic field,
stellar rotation rates, and disk resistivities. To illustrate the
trends, we plot in Figs.~\ref{jmes}-\ref{jtot}  the functions with different coefficients,
using triangles,
circles and squares, while approximately following the lines obtained from simulations. If
one function does not match all the cases, the lines indicate
different functions.

As an example, the goodness of fit shown in the top leftmost panel in the solutions with $\alpha_{\mathrm m}=0.1$ in Fig.~\ref{jmes},
measured by the R-squared, is 0.906 and 0.960 for
$\dot{J}_{\mathrm out}(\Omega_\star)/J_\star$ expressed as
$-130\Omega_\star^3$ and $-350\Omega_\star^3$ in the cases with
$\Omega_\star/\Omega_{\mathrm br}$ equal to 0.15 and 0.20,
respectively. In the bottom leftmost panel, goodness of fit, measured by the R-squared is 0.930 and 0.992 for
$\dot{J}_{\mathrm out}(B_\star)/J_\star$ expressed as $-0.8B_\star^3$
and $-2.7B_\star^3$ in the cases with $\Omega_\star/\Omega_{\mathrm br}$
equal to 0.15 and 0.20, respectively\footnote{We provide goodness of fits in this example for completeness and to illustrate that our
choice of expressions is well motivated. Still, it is a fit to only a four points, which is enough to establish trends, but should not be overstated for the functional dependencies.}. The
magnetospheric ejection expressions here offer the best fits, for the
other components the fit goodness is often less: in general, with
faster rotating star and larger magnetic field in which an accretion
column is not formed, the solution departs from the trend, because of
the change of flow geometry in the star-disk interaction system.

A division of the solutions in sub-classes with regard to the truncation
radius or disk mass accretion rate could offer better fits, as discussed
in \S2 of \cite{GalZanAma19}, to the relation to variable coefficients
of proportionality in the expressions for the torques by disk accretion
and magnetospheric ejections.

In the case of different mass fluxes in the various components in the flow, we computed the outcome in our simulations, which implicitly contains the information about the change in mass fluxes, as illustrated in
Figs.~\ref{mdotoutsfits} and \ref{mdotsom020} in the cases with outflow ($\alpha_{\mathrm m}=0.1$) and with the dependence on $\alpha_{\mathrm m}$, respectively. The comparison is more reliable between the cases where the mass fluxes in the components do not vary much because of the mixing of the flows measured at the different distances and instabilities. An example of a dependence of mass flux on the parameters included in the simulation is given in the rightmost panel in Fig.~\ref{mdotsom020}; this pertains to the case with 0.75 kG, where the disk mass accretion rate differs for a factor of 10 between the smallest and largest values of $\alpha_{\mathrm m}$ (0.1 and 1.0). The mass fluxes in other components of the flow also change.

As shown in Table~\ref{params}, in about a third of our cases (those
which have the names marked in bold letters), the total torque is
positive, $\dot{J}_{\mathrm tot}>0$, meaning that the central object
is spinning up. In general, the trend is that with larger
stellar magnetic field and faster stellar rotation (from the top left
towards the bottom right in the table), there are more spin-down
cases in our simulations. Outliers from this trend, such as simulations
a8, a12, c3, and d3, are often less stable cases, whereby the choice of
averaging interval could influence the trend. Another possibility is
that the trend is not valid for a smaller proportion of cases, for instance,
because of the different mass fluxes involved for different parameters.
Trends found in our simulations can be written with simple expressions,
which could be compared with the results from other models, simulations,
or observations, such as those in \cite{Ahuir20,Pantolmos20,GalZanAma19}.

For instance, the spin-up or spin-down timescale associated to the torques evaluated in our
various simulations is on the order or Myr. If we compare this result
with \cite{Gallet13} , in particular,  their Fig.~3, as well as the stellar rotation rates
in our sample (which cover the slow and median rotating stars of their
sample), we find that our results are in the correct range for solar-type
young stars.

\subsection{Stellar wind}
The torque by stellar wind, $\dot{J}_{\mathrm SW}$, in our simulations
is shown in Figs.~\ref{jswsa} and \ref{jsws}. The $\dot{J}_{\mathrm SW}$ is increasingly
negative with the increase in stellar rotation rate and stellar field
strength. In cases with magnetospheric ejection (left),
because part of the wind flow diverted into the magnetospheric
ejection, the increase is lower than in the cases without the ejection
(middle); it is also visible in the torques for the fastest
rotating stars in our sample (right). In Fig.~\ref{jswofalpha},
we show the dependence of stellar wind torque on stellar magnetic field
strength in the case of a viscous coefficient, $\alpha_{\mathrm v}=0.7,$ for
different stellar rotation rates. With the increasing stellar rotation rate,
the $\dot{J}_{\mathrm SW}/J_\star$ is increasingly negative, with the $B_\star^3$
dependence in the cases with larger magnetic fields.

The stellar wind dependence on the Alfv\'{e}n radius $r_{\mathrm A}$
from our simulations can be compared with Eq.~10 from \cite{GalZanAma19}:
$\dot{J}_{\mathrm SW}\propto\Omega_\star\dot{M}_{\mathrm SW}r_{\mathrm A}^2$. This can be
 combined with their Eq.~11 for the average Alfv\'{e}n radius to express:
\begin{equation}
r_{\mathrm A}=K_1R_\star\left(\frac{B_\star^2R_\star^2}{\dot{M}_{\mathrm SW}
\sqrt{K_2^2 v_{\mathrm esc}^2+\Omega_\star^2R_\star^2}}\right)^{m},
\label{ravalf}
\end{equation}
which gives:
\begin{equation}
\frac{dJ_{\mathrm SW}}{dt}=-K_1^2\Omega_\star R_\star^2\dot{M}_{\mathrm SW}
\left(\frac{B_\star^2R_\star^2}{\dot{M}_{\mathrm SW}\sqrt{K_2^2 v_{\mathrm esc}^2+
\Omega_\star^2R_\star^2}}\right)^{2m}.
\label{eq4GB}
\end{equation}  
Here $K_1=1.7$, $K_2=0.0506$ and $m=0.2177$ are determined from numerical
simulations of a stellar wind following the open field lines of a stellar
dipole \citep{Matt12a} and $v_{\mathrm esc}=\sqrt{2GM_\star/R_\star}$ is
the escape velocity. The stellar magnetic field is measured at the
stellar equator. In the top panels in Fig.~\ref{jcomppGB19}, we show
comparison of our results with the numbers obtained from Eq.~\ref{eq4GB}
(normalized to our units of $J_\star$). In the $\dot{J}_{\mathrm SW}$ cases
(shown in the middle panels), we obtained in the simulations the same direction
of the trend gradient as predicted, and our results are in agreement with
the prediction within an order of magnitude\footnote{In \cite{Gallet13}
the $K_1=1.3$ in Eq.~\ref{eq4GB} was given, but then the percentage of
the mass flux from the disk diverted into the stellar wind was assumed to
be 3\%, instead of 1\% assumed in \cite{GalZanAma19}.}. The constant
factor of 3 to 5 between their results and our simulations can
easily be overcome by taking into account the change in factor $K_1$
and adjusting the different mass fluxes, as these authors predicted. In the results shown
in the middle bottom panel in the same figure, we multiply $K_1$ by 2.35
to obtain a much better agreement.

The relations for $\dot{J}_{\mathrm SW}$ reported in \cite{Gallet13} and
\cite{GalZanAma19} have been confirmed by observations \citep{Gallet15}.

\subsection{Cases with magnetospheric ejection}
Flows and associated magnetic configurations in star-disk system are the
most complicated in the cases with $\alpha_{\mathrm m}=0.1$, when
a magnetospheric ejection is launched from the magnetosphere. It carries
away part of the material from the magnetosphere, together with its
angular momentum.

The torques exerted on the star by a magnetospheric interaction
with such ejection, $\dot{J}_{\mathrm out}$, and by the material
infalling from the disk onto the star, $\dot{J}_\star$, are shown in
Fig.~\ref{jmes}, together with total torques in the system. The same
results are shown in relations to different quantities, to reveal trends
in the solutions.

We find that torque on the star exerted by the magnetospheric ejection,
which slows down the star, increases with the fourth power of
rotation rate, as shown in the left top and bottom panels in
Fig.~\ref{jmes} (taking into account the $J_\star\propto\Omega_\star$
dependence from \S 4): $\dot{J}_{\mathrm out}(\Omega_\star)\propto(-\Omega_\star^4)$,
and with
$\dot{J}_{\mathrm out}(\Omega_\star, B_\star)\propto(-\Omega_\star B_\star^3)$.
The goodness of such fits is high ($R^2=0.992$) for the larger rotation
rates and stellar magnetic fields, for instance,  the more exact least-squares
fits to the polynomial $a_1B_\star^3+a_2B_\star^2+a_3B_\star+a_4$ would
give only a slightly better fit of the order $R^2=0.992$. For weaker
magnetic fields of 0.25 and 0.5 kG and more slowly rotating stars with 0.05
and 0.1 $\Omega_\star/\Omega_{\mathrm br}$, the dependences on the stellar
rotation rates and stellar magnetic field are weak and rather linear.

The trend in $\dot{J}_{\mathrm out}$ is increasing in both cases: with the
larger magnetic field and faster stellar rotation, the negative torque on
a star is increasing and the star is slowing down faster. The matter
load in the magnetospheric ejection is typically two to four orders of
magnitude smaller than the inflow onto the star. In the case of YSOs,
it yields values in the range of $10^{-11}-10^{-13} M_\sun yr^{-1}$.

Torques exerted on a star by material in-falling onto it through the
accretion column are shown in the middle panels in Fig.~\ref{jmes}. With
the different field strengths, shown in the top middle panel, the torque
does not change the stellar rotation rate:
$\dot{J}_\star/J_\star\propto 1/\Omega_\star\Rightarrow\dot{J}_\star=const$.
In the bottom middle panel, where the same results are organized by the
different stellar rotation rates, we note only a weak dependence on the
$\Omega_\star B_\star^3$. The results in this panel can be related to the stellar
field to match the results for other resistivities from
Fig.~\ref{jstar2}. The obtained proportionality to $\Omega_\star B_\star^3$
gives a good agreement for the faster rotating stars, while for more slowly
rotating stars, the dependence is not compelling. Still, there is a
clear trend in decrease of $\dot{J}_\star$ with faster stellar
rotation: a faster rotating star is less slowed down by the infalling
material\footnote{Note: for $\dot{J}_\star$ normalized to the
$J_\star\propto\Omega_\star$, for a faster rotating star, the
non-normalized value of $\dot{J}_\star$ will be larger, increasing the
spin-down or spin-up of the star.}. With the larger magnetic field and faster rotation, the mass load in the outflow is larger, as shown in Fig.~\ref{mdotoutsfits}. A larger centrifugal force, because of the larger mass in the outflow, would exert a larger torque. This could contribute to the $\dot{J}_{\mathrm out}\propto\Omega_\star^4$ dependence.

We compare our result with \cite{GalZanAma19}, adopting their Eq.~9:
\begin{equation}
\frac{dJ_{\mathrm out}}{dt}=K_{\mathrm ME}\frac{B_{\mathrm dip}^2R_\star^6}{R_{\mathrm
t}^3}\left[K_{\mathrm rot}-\left(\frac{R_{\mathrm t}}{R_{\mathrm
cor}}\right)^{3/2}\right].
\label{eq9GZA}
\end{equation}
Our results (shown in the right top panel in Fig.~\ref{jcomppGB19}), do
not match their prediction well for the larger stellar fields. For smaller
magnetic field strengths and stellar rotation rates, the match is improved.
The difference stems from the constants $K_{\mathrm ME}$ and $K_{\mathrm rot}$,
which would change with mass accretion rate and the ratio of truncation and
corotation radius (see the discussion in Sect.~2.3.4 in \cite{GalZanAma19}).
We show, in the bottom right panel in the same Fig.~\ref{jcomppGB19}, the result with changed
$K_{\mathrm ME}$ and $K_{\mathrm rot}$ (multiplied with factors 0.05 and 2.1,
respectively), which improves the outcome, for instance, in the case of
$\Omega_\star/\Omega_{\mathrm br}=0.15$. Similar tuning could be done in each of the
cases, but this is not our task here: in our simulations, the mass fluxes and radii
are matching self-consistently, including the varying mass fluxes in different
parts of the flow (as shown in Fig.~\ref{mdotsom020}).

The total torque in the system $\dot{J}_{\rm tot}/J_\star$, with
$\alpha_{\mathrm m}=0.1$, is shown in the rightmost panels in
Fig.~\ref{jmes}. It is following the same proportionality to
$B_\star^3$ and $\Omega_\star^3$. Since $J_\star\propto\Omega_\star$,
we have $\dot{J}_{\rm tot}\propto\Omega_\star B_\star^3$ and 
$\dot{J}_{\rm tot}\propto\Omega_\star^4$.

From the amounts of torque in the components, we see that magnetospheric
ejection dominates in the net torque of the system. The critical value of
stellar rotation rate at which the switch from a spin-up to a spin-down occurs
is about $\Omega_\star=0.1\Omega_{\mathrm br}$.

\subsection{Cases without magnetospheric ejection}
Our results with a resistive coefficient of $\alpha_{\mathrm m}>0.1$ do not
show any magnetospheric
ejection, resulting in a simpler flow pattern and field
configuration (shown in the top panel in Fig.~\ref{fig:sols1}).

Torques exerted on a star by infalling material from the disk
through an accretion column are shown in Fig.~\ref{jstar2}. The
dependence on the stellar rotation rate normalized to $J_\star$ is, again,
$\dot{J}_\star(\Omega_\star) =const$ and on the magnetic
field, it is $\dot{J}_\star(B_\star)\propto B_\star^{3}$;  here, it is
also $\dot{J}_\star(\Omega_\star, B_\star)\propto \Omega_\star B_\star^{3}$.

We checked that similar trends are followed by
$\dot{J}_{\mathrm R<R_{\mathrm cor}}$, which is the leading term in the
sum making the $\dot{J}_\star$. The contribution from
$\dot{J}_{\mathrm R>R_{\mathrm cor}}$ is in most cases an order of
magnitude smaller than $\dot{J}_{\mathrm R<R_{\mathrm cor}}$.

We again compare our results with the \cite{GalZanAma19} estimate. The
torque exerted on the star by the material inflowing through the
accretion column can be estimated by Eqs.~4-5 from \cite{GalZanAma19}
($\dot{J}_{\mathrm acc}$ in their notation):
\begin{equation}
\dot{J}_{\mathrm{acc}}=\frac{dJ_\star}{dt}=K_{\mathrm acc}\dot{M}_{\mathrm acc}\sqrt{GMR_{\mathrm
t}}\ ,
\label{eq4GZA}
\end{equation}
with
\begin{equation}
R_{\mathrm t}=K_{\mathrm t}
\left(\frac{B_{\mathrm dip}^4R_\star^{12}}{GM_\star\dot{M}_{\mathrm
acc}^2}\right)^{1/7},
\label{eq5GZA}
\end{equation}
from \cite{Bess08} and with $K_{\mathrm acc}=1$ for the cases without
magnetospheric ejections. The comparison with the results in our simulations
is shown in Fig.~\ref{jcomppGB19}. Our values match their prediction only at some values; however, for many, the discrepancy between the results from our simulations and their prediction is larger, and the action of torque is also different. Instead of speeding the star up, in our simulations, it is slowing it down.

In the bottom left panel of Fig.~\ref{jcomppGB19}, we show the same results from our simulations, but with the computation of torques (shown with dashed lines) performed by multiplying the factors $K_{\mathrm acc}$ with 1.1, 0.4, -0.5, and -2.5, respectively, for the increasing stellar field strengths, to better match the results from our simulations. This would amount to changes in the mass accretion rate with different field strengths. Again, in our simulations, this is done self-consistently.

All the components shown above contribute to the total torque exerted
on the star, shown in Fig.~\ref{jtot} for the $\alpha_{\mathrm m}=$0.4,
0.7 and 1 (left to right panels, respectively). It is the sum of components
from the different parts of flow and field configurations in the system.

In the cases with fastest stellar rotation in our study,
$\Omega_\star$=0.2, the total torque in the system is slightly out of
the trend. We present such cases in Fig.~\ref{jtotomalph}, also adding
 the cases with $\alpha_{\mathrm m}=0.1$, to show
that they follow a trend with the increasing magnetic field
strengths. In the cases with slower stellar rotation or smaller stellar
field, the total torque is often positive and its variation less steep.
Another result that can be appreciated from this figure (and this is another reason
why we added also the $\alpha_{\mathrm m}=0.1$ cases) is that the total
torque $\dot{J}_{\mathrm tot}$ does not depend much on the resistive
coefficient, $\alpha_{\mathrm m}$. The reason for departing from the
trend in the cases with faster stellar rotation is not the resistivity
in the disk, but the different geometry of the system: the disk is
pushed away from the star and there is no accretion column. Most of the
torque on the star now comes from the part of the system beyond the
corotation radius, $R_{\mathrm cor}$, slowing the star down. In the bottom
panel in the same figure we show the dependence of total torque on
the stellar rotation rates with $\alpha_{\mathrm m}$ for the simulations
with B$_\star=0.5$~kG. Here, we also see weak dependence on
$\alpha_{\mathrm m}$.

In Fig.~\ref{jtotoom}, we show the dependence of total torque in the
system (normalized to the $J_\star$) with the stellar rotation rate,
$\Omega_\star$, and its trend with the magnetic field, $B_\star$, in
cases with $\alpha_{\mathrm m}=1$. When we include the
$J_\star\propto\Omega_\star$ dependence, we find that
$\dot{J}_{\mathrm tot}(\Omega_\star)=const$. Similar results are
obtained with smaller $\alpha_{\mathrm m}$, namely, in most of the
cases, we obtain a spin-down for the central object. Here, we also
see that the critical stellar rotation rate for a switch from spin-up to
spin-down is between 0.07 and 0.11 $\Omega_{\mathrm br}$, as
in the cases with magnetospheric ejection.

In most of the cases without magnetospheric ejection, the magnetic field
is anchored in the disk beyond the corotation radius, with the
accretion column positioned below the corotation radius (as in
simulation b8, illustrated in Fig.~\ref{fig:sols1}). With the
increase of stellar rotation rate, corotation radius shifts closer to
the star, approaching the footpoint of the accretion column. This, in
turn, can change the torque, and cause a switch between the
spin-up and spin-down of the star. When the central object is slowed-down,
the decrease in the stellar rotation is slower than in the cases with
the magnetospheric ejection with the stronger field.

\section{Conclusions}\label{conclu}
In our numerical simulations of a star-disk magnetospheric interaction,
we obtained a suite of quasi-stationary solutions. We performed a parameter
study based on 64 axisymmetric 2D MHD simulations, varying the disk
resistivity, stellar dipole magnetic field strength, and rotation rate.

In order to assess how far the stellar magnetic field is able to
connect itself into the disk, we measured the furthest anchoring radius,
$R_{\mathrm out}$. In our simulations, we find the following trends:\\
$\bullet$ $R_{\mathrm out}$ increases with the larger resistive
coefficient, $\alpha_{\mathrm m}$, for all the strengths of the stellar
magnetic field. This is because the field line is able to slip
through the disk more easily than in less diffusive case, where
it disconnects due to strong shear. \\

$\bullet$ In cases with $\alpha_{\mathrm m}=0.1$, when a magnetospheric
ejection is launched from the system, there is only a minor dependence
of $R_{\mathrm out}$ on the stellar rotation rates for all the stellar
field strengths. The increase in $R_{\mathrm out}$ with the stellar
field strength is small in such cases. This is likely due to the field
geometry and the presence of the current sheet at mid latitudes. 

In all the cases, we find that the kinematic torque is negligible,
implying that most of the torque comes from the magnetic interaction.

We describe the dependence of torques in the system by characterizing
their magnetospheric interaction regime with approximate expressions.
We obtain:\\
$\bullet$ The torque exerted on the star by a material in-falling from
the disk onto the star, $\dot{J}_\star$, by a material expelled from
the system in a magnetospheric ejection,
$\dot{J}_{\mathrm out}$, and a total torque in the system, $\dot{J}_{\mathrm tot}$,
we can write in all the cases as:
\begin{eqnarray}
\dot{J}_\star,\ \dot{J}_{\mathrm out},\ \dot{J}_{\mathrm tot} \propto
\Omega_\star B_\star^3. \nonumber
\end{eqnarray}

$\bullet$ In all the cases, the total torque in the system does not
depend much on the resistivity coefficient, $\alpha_{\mathrm m}$.

$\bullet$ Our results for stellar wind are in a reasonable agreement with the theoretical and observational results on magnetospheric star-disk interaction and stellar wind from \cite{GalZanAma19},
with trends in $\dot{J}_{\mathrm SW}$ following the predicted expression (see Fig.~\ref{jcomppGB19}). We show that their assumption of constant factor, K$_1$, is not in
agreement with the results we obtained with the self-consistent treatment in MHD simulations.\\
$\bullet$ In all the cases without magnetospheric ejection, the torque
exerted on the star is independent of stellar rotation rate:
\begin{eqnarray}
\dot{J}_\star(\Omega_\star)=const.\nonumber
\end{eqnarray}
In our simulations, these are all the cases with $\alpha_{\mathrm m}>0.1$. Here, we also show that assumption of the constant, K$_{\mathrm acc}$, from
\cite{GalZanAma19} is not in agreement with our self-consistent treatment in simulations.

$\bullet$ In all the cases with $\alpha_{\mathrm m}=0.1$, abcd(1,5,9,13), a
magnetospheric ejection is launched in our simulations. Most of the torque in the
system in such cases is in the ejection. The component of the torque exerted
on the star by such an ejection can be expressed as:
\begin{eqnarray}
\dot{J}_{\mathrm out}(\Omega_\star)\propto \Omega_\star^4. \nonumber
\end{eqnarray}
From our results we conclude that the reason for such strong dependence
is the fact that the faster rotating star increases the amount of material in the outflow, which results in a larger torque and centrifugal force.\\
$\bullet$ In two-thirds of all the cases with magnetospheric ejection in our
simulations, the central star is spun up. The spin-up stops in the cases with
larger field and faster stellar rotation. In the cases without magnetospheric
ejection, only a third are yielding a spun-up star. With the increasing
stellar magnetic field or faster stellar rotation rate, we observe a switch
of sign in the net torque, resulting in a spun-down star. The spin-down is
also increasing with the increasing field strength or stellar rotation rate.\\
$\bullet$ The critical stellar rotation rate at which the spin-up switches to spin-down is between 0.07 and 0.11 $\Omega_{\mathrm br}$.\\
$\bullet$ A comparison with \cite{GalZanAma19} results shows that the
constant factors K$_{\mathrm acc}$ and K$_{\mathrm ME}$ from their expressions are not in agreement with the self-consistent treatment in our simulations, we instead find a variation among these prefactors. It is a consequence of changes in the mass fluxes in the different components in the flow with the different stellar rotation rates, magnetic field, and disk resistivity. For example, in our simulations, we find that in cases with magnetospheric ejections ($\alpha_{\mathrm m}=0.1$), the mass losses through the stellar wind and outflow $\dot{M}_{\mathrm out}$ and $\dot{M}_{\mathrm SW}$ are both proportional to $\Omega_\star^6$ and $B_\star^4$, and the mass fluxes $\dot{M}_\star$ onto the star are proportional to $\Omega_\star^3$. In the cases with other $\alpha_{\mathrm m}$, scattering in the results is larger.\\

Here, we list the most important caveats in our work. Our sample of numerical
simulations was limited to slowly\footnote{In the designation used in \cite{Gallet13}, our sample includes slow and median rotating stars.} rotating
objects at up to 20\% of the breakup velocity at the equator. 
For the faster rotating objects, an axial outflow often forms,
which would further complicate the description. We started
a separate line of study for such cases \citep{Kotek20}, where a more
thorough study of torque in magnetospheric ejections will, in connection
with the axial outflow, be more complete. Mass fluxes in the different components of the flow demand a separate study, probably with an additional division of the results with a positioning of the characteristic radii in the system.  Also, we refer here only to stellar dipole fields, when it is known that multipole stellar fields are closer to reality-this is another separate line in our research \citep{Cieciuch22}. Another complication we avoided by setting the viscous
anomalous coefficient
$\alpha_{\mathrm v}=1$ in all simulations to avoid backflow in the disk.
Such an outflow near the disk midplane, directed away from the star was also
found in numerical computations with alpha-viscosity
\citep{Kley92, Igum95, Roz94} and with the magneto-rotational instability
\citep{White20, MishraB20}. We describe it elsewhere
\citep{MishraR20a,MishraR20b,MishraR22}. In our computation of torques, we
checked that the field in most of the disk is at least one order of magnitude
smaller than the stellar field, so we neglected the effect of magnetic field
inside the accretion disk on the result. However, \cite{LN13} showed that in
some cases the disk field affects the torque on the star; in our solutions, it
would be in the cases when corotation radius is nearby the footpoint of the
accretion column. We leave this point for  a future study.

\section*{Acknowledgements}
 M\v{C} acknowledges the Czech Science Foundation (GA\v{C}R) grant No.~21-06825X
 and the Polish NCN grant 2019/33/B/STA9/01564. M\v{C} developed the setup for
star-disk simulations while in CEA, Saclay, under the ANR Toupies grant,
and a collaboration with the Croatian STARDUST project through HRZZ
grant IP-2014-09-8656 is acknowledged. M\v{C} thanks to the support by the
International Space Science Institute (ISSI) in Bern, which hosted the
International Team project \#495 (Feeding the spinning top) with its inspiring
discussions. A.S. Brun acknowledges support by CNES PLATO grant and ERC Stars~2.
We thank IDRIS (Turing cluster) in Orsay, France, ASIAA (PL and XL clusters) in
Taipei, Taiwan and NCAC (PSK and CHUCK clusters) in Warsaw, Poland, for access
to Linux computer clusters used for the high-performance computations. The
{\sc PLUTO} team, in particular A. Mignone, is thanked for the possibility to
use the code.


\bibliographystyle{aa} % style aa.bst
\bibliography{angmomprob1finarx} % your references Yourfile.bib

%%%%%%%%%%%%%%%%% APPENDICES %%%%%%%%%%%%%%%%%%%%%

%\appendix
\nolinenumbers
\end{document}